\documentclass[final,5p,times,twocolumn]{elsarticle}

\usepackage{amssymb}

\usepackage{color}
\usepackage{xcolor}
\usepackage{subfigure}
\usepackage[hidelinks]{hyperref}
\usepackage[printonlyused]{acronym}
\usepackage{listings}
\usepackage{algorithm,algpseudocode}
\usepackage[rflt]{floatflt}
\usepackage{stfloats}

\newcommand{\bq}{\begin{equation}}
\newcommand{\eq}{\end{equation}}

\newcommand{\byte}{\mbox{byte}}
\newcommand{\second}{\mbox{s}}
\newcommand{\seconds}{\mbox{s}}
\newcommand{\flop}{\mbox{flop}}

\newcommand{\GBS}{\mbox{G\byte/\second}}

\newcommand{\GFS}{\mbox{G\flop/\second}}

\newcommand{\GHZ}{\mbox{GHz}}

\newcommand{\GB}{\mbox{GB}}

\newcommand{\MiB}{\mbox{MiB}}
\newcommand{\KiB}{\mbox{KiB}}

\newcommand{\muops}{\mbox{$\mu$-ops}}

\definecolor{tumbleweed}{rgb}{0.87, 0.67, 0.53}

\def\addlegendimage{\csname pgfplots@addlegendimage\endcsname}

\definecolor{applegreen}{rgb}{0.55, 0.71, 0.0}
\definecolor{amethyst}{rgb}{0.6, 0.4, 0.8}
\definecolor{amber}{rgb}{1.0, 0.75, 0.0}

\journal{Future Generation Computer Systems}

\newacro{AoS}{array of structures}
\newacro{AoSoA}{array of structures of arrays}
\newacro{CPI}{cycles per instruction}
\newacro{DEM}{discrete element method}
\newacro{EAM}{embedded atom method}
\newacro{FCC}{face-centered cubic}
\newacro{FN}{full neighbor-lists}
\newacro{HN}{half neighbor-lists}
\newacro{ILP}{instruction-level parallelism}
\newacro{IR}{intermediate representation}
\newacro{ISA}{instruction set architecture}
\newacro{LJ}{Lennard-Jones}
\newacro{MD}{Molecular dynamics}
\newacro{MPI}{message passing interface}
\newacro{SIMD}{single instruction, multiple data}
\newacro{SoA}{structure of arrays}
\newacro{PBC}{periodic boundary conditions}
\newacro{HPM}{hardware performance monitoring}

\begin{document}

\begin{frontmatter}



\title{MD-Bench: Engineering the in-core performance of short-range molecular dynamics kernels from state-of-the-art simulation packages}


\author[nhratfau]{Rafael Ravedutti Lucio Machado}
\author[nhratfau]{Jan Eitzinger}
\author[nhratfau]{Jan Laukemann}
\author[nhratfau]{Georg Hager}
\author[nhratfau]{Harald Köstler}
\author[nhratfau]{Gerhard Wellein}

\address[nhratfau]{Erlangen National High Performance Computing Center, Martensstraße 1, Erlangen, 91058, Bayern, Germany}

\begin{abstract}
Molecular dynamics (MD) simulations provide considerable benefits for the investigation and experimentation of systems at atomic level.
Their usage is widespread into several research fields, but their system size and timescale are also crucially limited by the computing power they can make use of.
Performance engineering of MD kernels is therefore important to understand their bottlenecks and point out possible improvements.
For that reason, we developed MD-Bench, a proxy-app for short-range MD kernels that implements state-of-the-art algorithms from multiple production applications such as LAMMPS and GROMACS.
MD-Bench is intended to have simpler, understandable and extensible source code, as well as to be transparent and suitable for teaching, benchmarking and researching MD algorithms.
In this paper we introduce MD-Bench, describe its design and structure and implemented algorithms.
Finally, we show five usage examples of MD-Bench and describe how these are useful to have a deeper understanding of MD kernels from a performance point of view, also exposing some interesting performance insights.
\end{abstract}

\begin{graphicalabstract}
\end{graphicalabstract}

\begin{highlights}
\item Molecular dynamics proxy-app for performance engineering
\item Systematic performance analysis with hardware performance counters
\item Evaluation of memory latency and control flow divergence contributions
\item Verlet lists and GROMACS MxN optimization kernels for short-range force calculation
\end{highlights}


\end{frontmatter}



\section{Introduction and Motivation}\label{sec:introduction}
\ac{MD} simulations are used in countless research efforts to assist the investigation and experimentation of systems at the atomic level in such diverse fields as materials science, engineering, natural science and life sciences.
The system size and timescale for MD simulations are crucially limited by computing power, therefore they must be designed with performance in mind.
Several strategies to speed up such simulations exist, examples are Linked Cells, Verlet List and MxN kernels from GROMACS \cite{PALL20132641,DBLP:journals/jcc/SpoelLHGMB05}.
These improve the performance by exploiting either domain knowledge or hardware features like SIMD capabilities and GPU accelerators.
\ac{MD} application codes can achieve a large fraction of the theoretical peak floating-point performance and are therefore among the few application classes that can make use of the available compute power of modern processor architectures.
Also, cases scientists are interested in are frequently strong scaling throughput problems; a single run only exhibits limited parallelism but thousands of similar jobs need to be executed.
This fact combined with arithmetic compute power limitation makes an optimal hardware-aware implementation a critical requirement.

Proxy-apps are stripped-down versions of real applications. Ideally they are self-contained, easy to build and bundled with a validated test case, which can be a single ``stubbed'' performance-critical kernel or a full self-contained small version of an application.
Historically, proxy-apps helped with porting efforts or hardware-software co-design studies.
Today, they are a common ingredient of any serious performance engineering effort of large-scale application codes, especially with multiple parties involved.
Typically, proxy-apps resemble a single application code; e.g., miniMD \cite{6805038} mimics the performance of LAMMPS \cite{PLIMPTON19951,BROWN2012449}.
A proxy-app can be used for teaching purposes and as a starting point for performance oriented-research in a specific application domain.

To investigate the performance of MD applications, we developed \textit{MD-Bench}, a standalone proxy-app toolbox implemented in C99 that comprises the most essential \ac{MD} steps to calculate trajectories in an atomic-scale system.
MD-Bench is intended to facilitate and encourage performance-related research for classical MD algorithms.
It contributes clean reference implementations of state-of-the-art MD optimization schemes.
As a result, and in contrast to existing MD proxy-apps, MD-Bench is not limited to representing one MD application but aims to cover all relevant contributions.
Its applications are low-level code analysis and performance investigation via fine-grained profiling of hardware resources utilized by \ac{MD} runtime-intensive kernels.

In previous work \cite{mdbench-ppam}, we presented the features of MD-Bench including some showcases. In this paper we extend the coverage as follows:
\begin{itemize}
    \item We added a systematic evaluation using hardware performance metrics for employing single-precision data types on \ac{MD} kernels for both LAMMPS and GROMACS algorithms (\autoref{sec:fpanalysis}).
    \item We make use of the most recent Intel C++ Compiler Classic (ICC) and Intel oneAPI DPC++/C++ Compiler (ICX). These resulted in better performance than all other compilers that were evaluated.
    \item We added an analysis for generated codes with new compilers and AVX2 variants (\autoref{sec:asm}). Such analysis includes the use of hardware gather instructions, a preamble variant for the innermost neighbor traversal loop with the widest SIMD instruction set available, the usage of reciprocal instead of a division instruction and the presence of Newton-Raphson iterations.
    \item We assess latency and control flow divergence contributions (\autoref{sec:latency}) for several new kernels and targets: (a) LAMMPS AVX512 on Ice Lake, (b) LAMMPS AVX2 on Cascade Lake and Ice Lake, (c) GROMACS AVX512 (both single- and double-precision cases) on Cascade Lake and Ice Lake, and (d) LAMMPS AVX2 on AMD Milan~(Zen3) system. Additionally, we include static analysis predictions from the LLVM Machine Code Analyzer (llvm-mca) tool and extend the predictions from OSACA (which only considers the backend) by a naive frontend analysis; the predictions are used as optimistic (lower-limit) runtime.
    \item We perform an additional \ac{HPM} analysis based on instruction counts to the Verlet List algorithm with \ac{FN} and \ac{AoS} data layout for the DP and SP versions to study how well different compilers can vectorize these cases using compiler flags to enforce AVX512 SIMD vectorization.
\end{itemize}
Note that our goal is not to carry out a compiler contest but rather analyze the performance of kernels that can be considered optimized and take advantage of SIMD features available in the target hardware.
GCC, CLANG and even the AMD Optimizing C/C++ (AOCC) were evaluated but in most cases did not generate competitive code in relation to ICC and ICX.


This paper is structured as follows:
In \autoref{sec:relatedwork} we present related work on proxy-apps for \ac{MD} simulations, pointing out the differences compared to MD-Bench.
In \autoref{sec:md} we explain the basic theory for \ac{MD} simulations.
In \autoref{sec:features} we list and discuss current features offered in MD-Bench, besides its employment on benchmarking, performance analysis and teaching activities.
In \autoref{sec:showcases} we present case studies to illustrate how MD-Bench can be used.
Finally, \autoref{sec:outlook} presents the conclusion and outlook, and a discussion of future work.

\section{Related Work}
\label{sec:relatedwork}
There already exist multiple proxy-apps to investigate performance and portability for \ac{MD} applications.
One of the better-known examples is \textit{Mantevo miniMD}, which contains C++ code extracted from LAMMPS and provides a homogeneous copper lattice test case in which the short-range forces can be calculated with \ac{LJ} or \ac{EAM} potentials.
It was used to investigate the performance of \ac{SIMD} vectorization on the innermost loops of the neighbor-lists building and force calculation steps~\cite{6569887}, as well as to evaluate the portability of \ac{MD} kernels through the Kokkos framework, thus executing most of the code on GPU instead of only the pair force and neighbor-lists.
Positive outcomes from the miniMD analysis in LAMMPS included better \ac{SIMD} parallelism on compiler-generated code and more efficient use of GPUs by avoiding data transfers in each time step.

\textit{ExMatEx coMD}~\cite{comd} is another proxy-app from the materials science domain that focuses on co-design to evaluate the performance of new architectures and programming models.
Besides allowing users to extend and/or re-implement the code as required, the co-design principle also permits to evaluate the performance when switching strategies, for example using Linked Cells for force calculations instead of Verlet Lists.

\textit{ExaMiniMD}~\cite{examinimd} is an improved and extended version of miniMD with enhanced modularity that also uses Kokkos for portability. Its main components such as force calculation, communication and neighbor-list construction are derived classes that access their functionality through virtual functions.

In previous work we developed \textit{tinyMD}~\cite{MACHADO2021101425}, a proxy-app (also based on miniMD) created to evaluate the portability of \ac{MD} applications with the AnyDSL framework.
TinyMD uses higher-order functions to abstract device iteration loops, data layouts and communication strategies, providing a domain-specific library to implement pair-wise interaction kernels that execute efficiently on multi-CPU and multi-GPU targets.
Its scope is beyond \ac{MD} applications, since it can be used for kind of particle simulation that relies on short-range force calculation such as the \ac{DEM}.

Beyond proxy-apps, performance engineering of \ac{MD} can be achieved via auto-tuning by either running simulations as a black-box for finding optimal system- and hardware-specific simulation parameters~\cite{doi:10.1063/5.0019045} or by providing programming interfaces that dynamically tune the application at runtime by selecting the best optimization strategies and data layouts~\cite{8778280}.

MD-Bench differs from all of these because it was primarily developed to enable an in-depth analysis of software-hardware interaction.
With that in mind, MD-Bench provides a ``stubbed'' variant to evaluate the force calculation at different levels of the memory hierarchy, as well as a gather benchmark that mimics the memory operations used in those kernels, thus allowing investigation of the memory transfers without side effects from arithmetic operations.
In contrast to other proxy-apps, MD-Bench contains optimized algorithms from multiple MD applications and allows to compare those using the same test cases.
Apart from standard Verlet Lists algorithms it contains state-of-the-art optimization strategies like, e.g., the GROMACS MxN kernels, which attain higher data-level parallelism on modern architectures by using a more SIMD-friendly data layout and which are not yet available as part of a simple proxy-app.
Although the significant majority of MD-Bench code is implemented in C, it also relies on SIMD intrinsics and assembly code kernels, which enables low-level tweaking of the code without interference from a compiler.



\section{Background and Theory}
\label{sec:md}

\ac{MD} simulations are widely used today to study the behavior of microscopic structures.
These simulations reproduce the interactions among atoms in these structures on a macroscopic scale while allowing us to observe the evolution in the system of particles in a time-scale that is simpler to analyze.

Different areas such as material science to study the evolution of the system for specific materials, chemistry to analyze the evolution of chemical processes, or biology to reproduce the behavior of certain proteins and bio-molecules resort to simulations of \ac{MD} systems.

A \ac{MD} system to be simulated constitutes of a number of atoms, the initial state (such as the atoms' position or velocities), and the boundary conditions of the system.
Here, we use \ac{PBC} in all directions, hence when one atom crosses the domain, it reappears on the opposite side with the same velocity.
Fundamentally, atom trajectories in classical \ac{MD} systems are computed by integrating Newton's second law for every atom $i$:
\begin{equation}
    \hat F_i = m \dot{\hat v}_i = m \hat a_i \label{eq:newton} 
\end{equation}
The forces $F_i$ are described by the negative gradient of the potential of interest.
The mutual force between atom $i$ and $j$ as governed by the \ac{LJ} potential
is given as
\begin{equation}
    \hat{F}_{2}^{LJ}(\hat{x_i}, \hat{x_j}) = -\frac{24\varepsilon}{x_{ij}} \left( \frac{\sigma}{x_{ij}} \right)^{6} \left[ 2\left(\frac{\sigma}{x_{ij}}\right)^{6} - 1\right] \hat{x}_{ij}~,
    \label{eq:lennard_jones}
\end{equation}
where $\hat{x}_{ij}$ is the distance vector between atoms $i$ and $j$, $x_{ij}$ is the distance between atoms $i$ and $j$, $\varepsilon$ is the width of the potential well, and $\sigma$ specifies at which distance the potential is~$0$.

\begin{figure*}[htb]
    \centering
    \subfigure[Verlet List]{\label{fig:verlet}\includegraphics[width=0.2\textwidth]{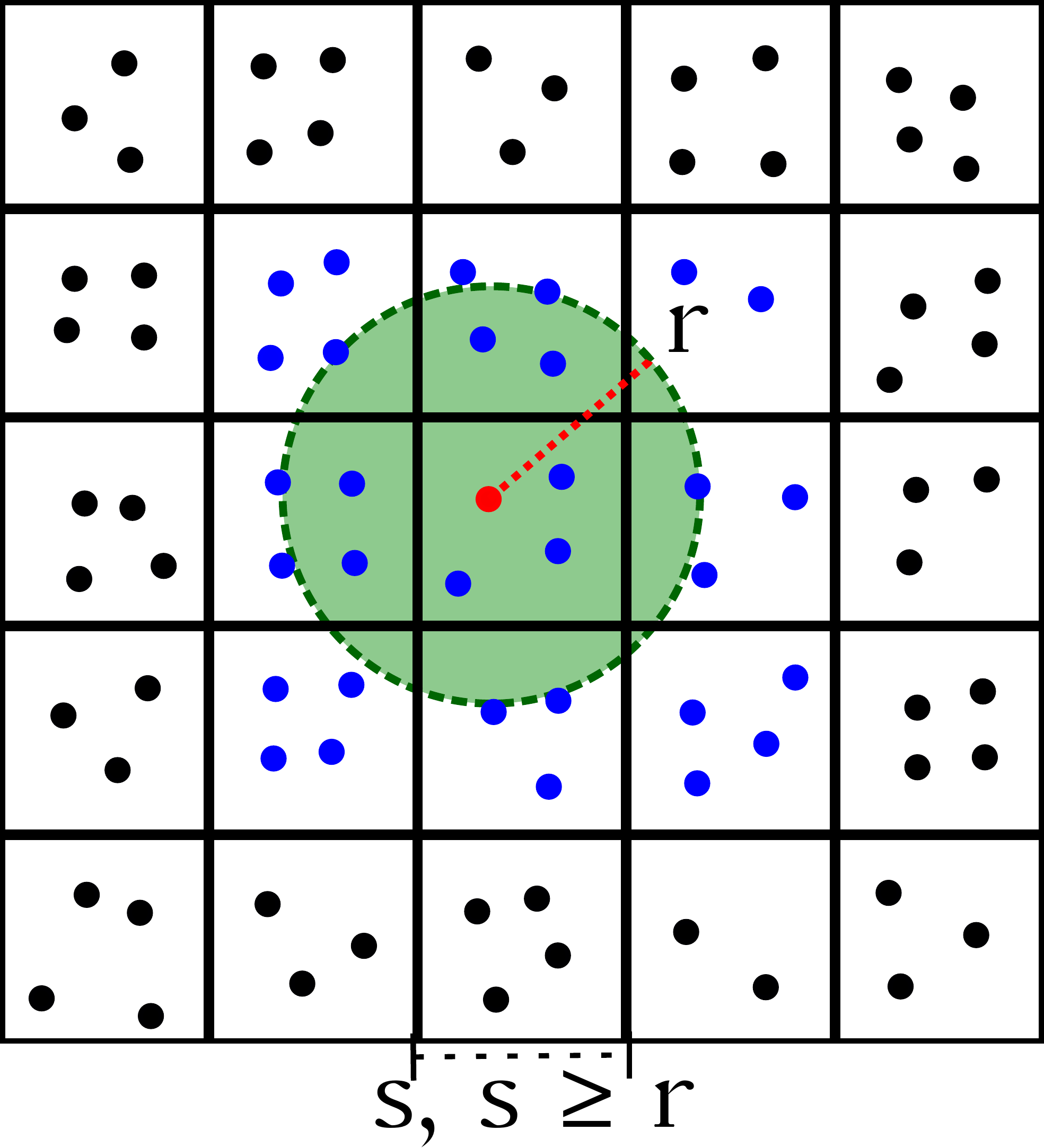}}
    \subfigure[GROMACS MxN]{\label{fig:gromacs_mxn}\includegraphics[width=0.2\textwidth]{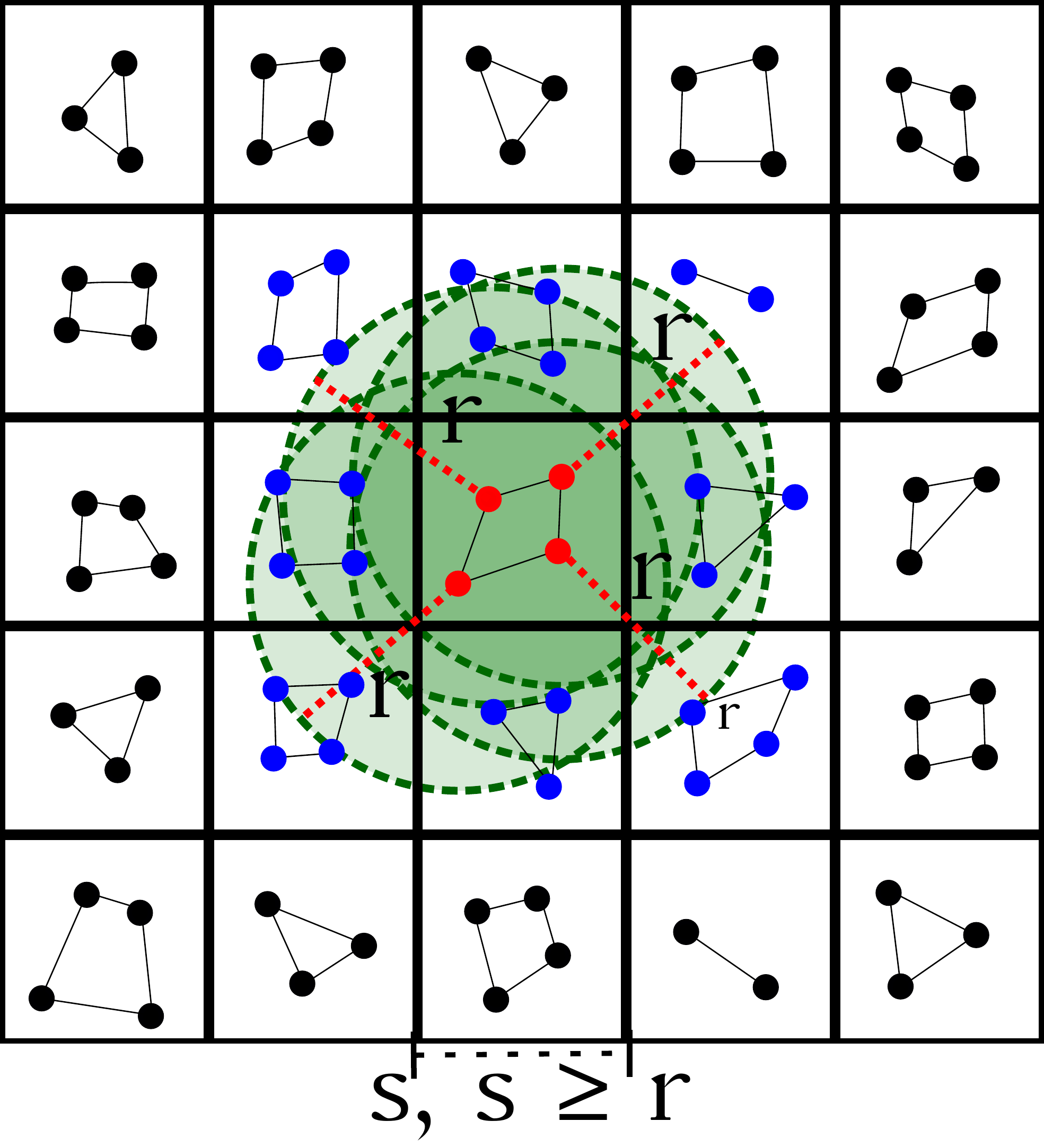}}
    \subfigure[Stubbed Patterns]{\label{fig:stub}\includegraphics[width=0.2\textwidth]{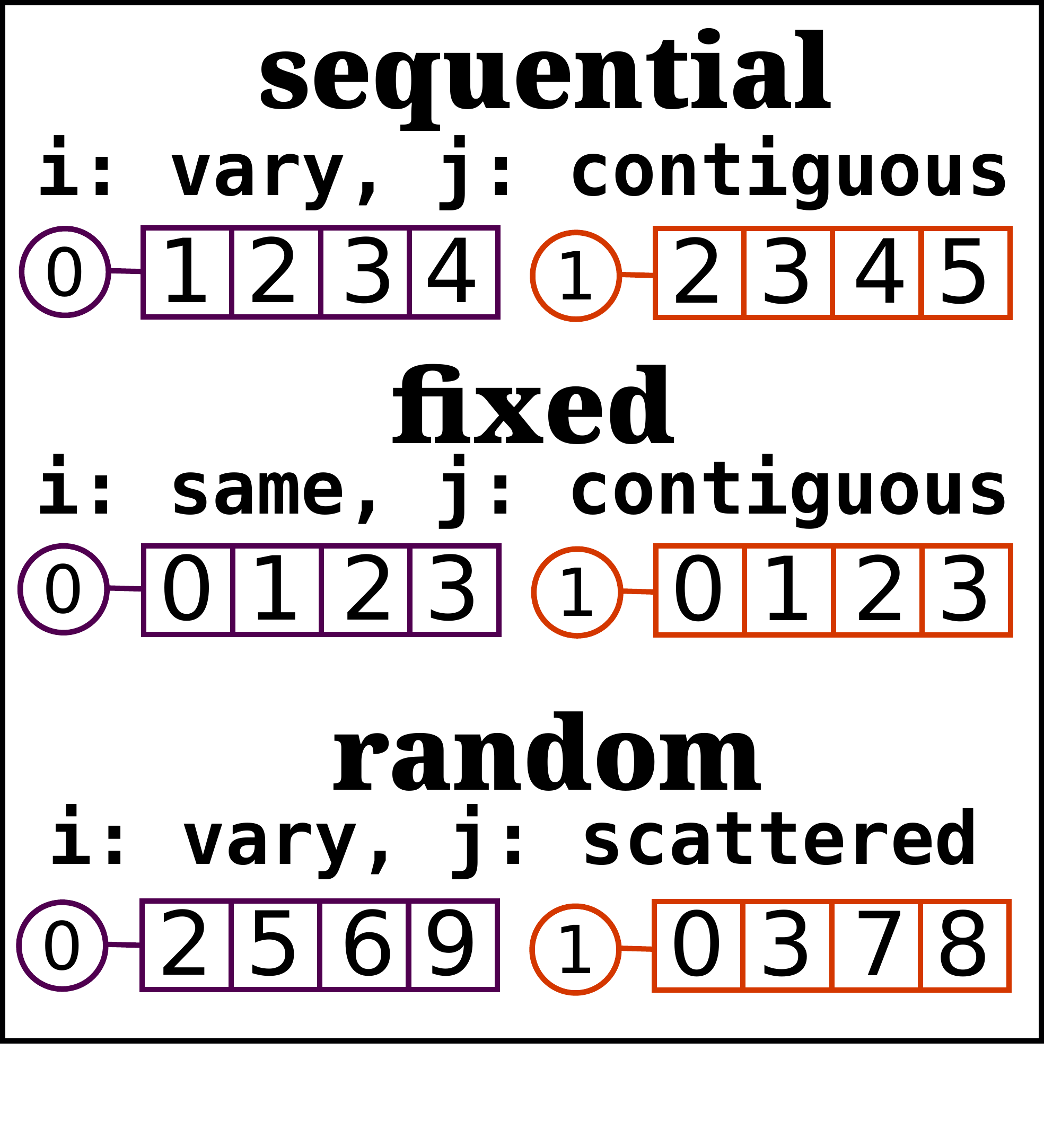}}
    \caption{(a) Pair list creation for the red atom in a Verlet List. Blue atoms are evaluated and the ones within the green circle with radius $r$ are added to the pair list.  (b) Pair list creation for the red atom cluster in GROMACS MxN, where the cutoff radius applies to each atom in the cluster. In both cases, the cell size $s$ must be greater or equal than $r$. (c) 4-length neighbor lists for atoms 0 (purple) and 1 (orange) in Stubbed Case patterns.}
\end{figure*}

To optimize the computation of short-range potentials, only atom pairs within a cutoff radius may be considered because contributions become negligible at long range distances.
First, cell lists are used to segregate atoms according to their spatial position.
As long as the cell sizes are equal to or greater than the cutoff radius, it is just necessary to search for neighbor candidates for an atom iterating over the cell it belongs to and the surrounding neighbor cells.
Thus, cell lists are used to build a Verlet list (see Figure~\ref{fig:verlet}) for each atom, tracking neighbor atoms within a specific radius $r$, which is the cutoff radius plus a small value (verlet buffer).
The verlet buffer extends the search range to avoid building the Verlet lists every time-step.

\section{MD-Bench Features}
\label{sec:features}

\begin{figure*}[htb]
    \centering
    \includegraphics[width=0.7\textwidth]{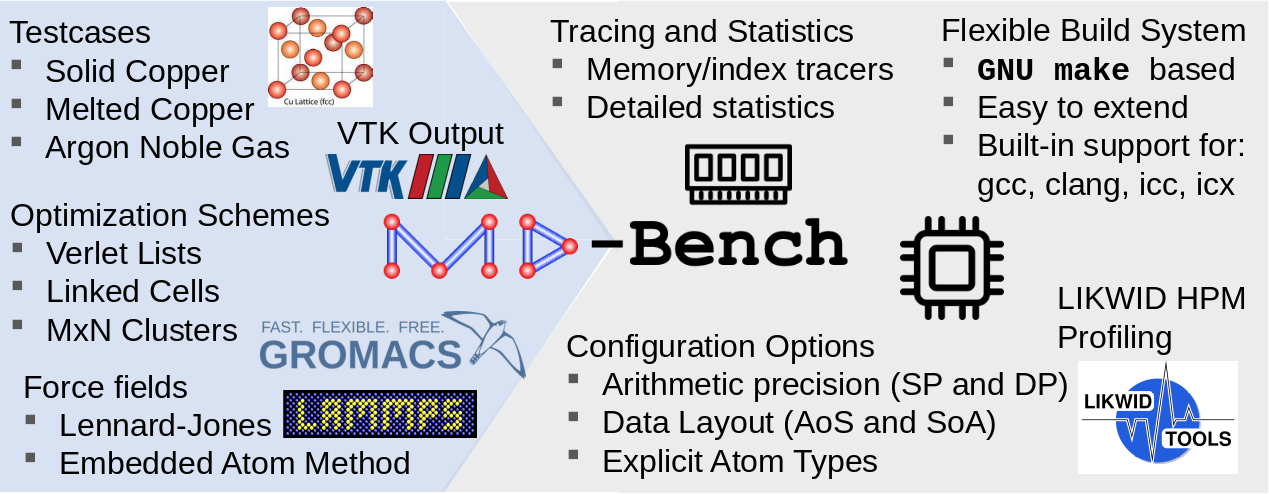}
    \caption{Overview of MD-Bench Features}
    \label{fig:features}
\end{figure*}


\autoref{fig:features} depicts MD-Bench\footnote{MD-Bench is open source and available under LGPL3 License at \url{https://github.com/RRZE-HPC/MD-Bench}.} features. To facilitate experimentation with a range of settings that influence performance, a robust build system with various configurations from the compiler and flags to whether atom types should be explicitly stored and loaded from memory is available.
Due to its modularity, the build system permits to replace kernels at the assembly level.
In particular, we maintain simplified C versions of each kernel which were written with special attention to low-level code analysis and tweaking.  This is hardly achievable on production \ac{MD} applications due to their massive code base and extensive use of advanced programming language techniques, which introduces substantial complexity and makes analysis more difficult.
Moreover, kernels are instrumented with LIKWID~\cite{psti} markers to allow fine-grained  profiling via \ac{HPM} counters.



\subsection{Optimization schemes}
\subsubsection{Verlet neighbor lists}

The most common optimization scheme used in \ac{MD} applications to compute short-range forces is arguably the Verlet List algorithm.
It consists of building a neighbor list for each atom, where the elements are other atoms that lie within a cutoff radius that is greater or equal than the force cutoff radius.
Thus, forces are computed for each local atom (i.e. atoms that belong to current process and are not \ac{PBC} shifted versions from other local atoms) by traversing its neighbor list and accumulating all forces for neighbors whose distance is smaller than the cutoff (see Algorithm~\autoref{alg:forcefn}).


\begin{algorithm}
\caption{Force calculation with a Verlet List (full neighbor-lists)}
\label{alg:forcefn}
\begin{algorithmic}
\Statex
    \scriptsize
    \For{$i \gets 1$ to $Nlocal$} \Comment{Number of local atoms}
    \State {$f_i$ $\gets$ {$0$}} \Comment{Partial forces (required for parallelism)}
    \For{$k \gets 1$ to $Nneighs[i]$} \Comment{Number of neighbors for atom $i$}
        \State {$j$ $\gets$ {$neighbors[i,k]$}} \Comment{$k$-th neighbor of atom $i$}
        \State {$d$ $\gets$ {$calculate\_distance(i, j)$}}
        \If {$d\leq cutoff\_radius$} \Comment{Force cutoff check}
            \State {$f_i$ $\gets$ {$f_i+calculate\_force(d)$}} \Comment{Depends on the potential used}
        \EndIf
    \EndFor
    \State {$force[i]$ $\gets$ $force[i] + f_i$} \Comment{Accumulate forces}
\EndFor
\end{algorithmic}
\end{algorithm}

\begin{algorithm}
\caption{Force calculation with a Verlet List (half neighbor-lists)}
\label{alg:forcehn}
\begin{algorithmic}
\Statex
    \scriptsize
    \For{$i \gets 1$ to $Nlocal$} \Comment{Number of local atoms}
    \State {$f_i$ $\gets$ {$0$}} \Comment{Partial forces (required for parallelism)}
    \For{$k \gets 1$ to $Nneighs[i]$} \Comment{Number of neighbors for atom $i$}
        \State {$j$ $\gets$ {$neighbors[i,k]$}} \Comment{$k$-th neighbor of atom $i$}
        \State {$d$ $\gets$ {$calculate\_distance(i, j)$}}
        \If {$d\leq cutoff\_radius$} \Comment{Force cutoff check}
	    \State {$f_{ij}$ $\gets$ {$calculate\_force(d)$}}
            \If {$j\leq Nlocal$} \Comment{Check if j is local atom}
		\State {$force[j]$ $\gets$ $force[j] - f_{ij}$} \Comment{Accumulate forces for j atom}
            \EndIf
	    \State {$f_i$ $\gets$ {$f_i+f_{ij}$}} \Comment{Depends on the potential used}
        \EndIf
    \EndFor
    \State {$force[i]$ $\gets$ $force[i] + f_i$} \Comment{Accumulate forces for i atom}
\EndFor
\end{algorithmic}
\end{algorithm}

\begin{figure}
    \centering
    \includegraphics[width=6cm]{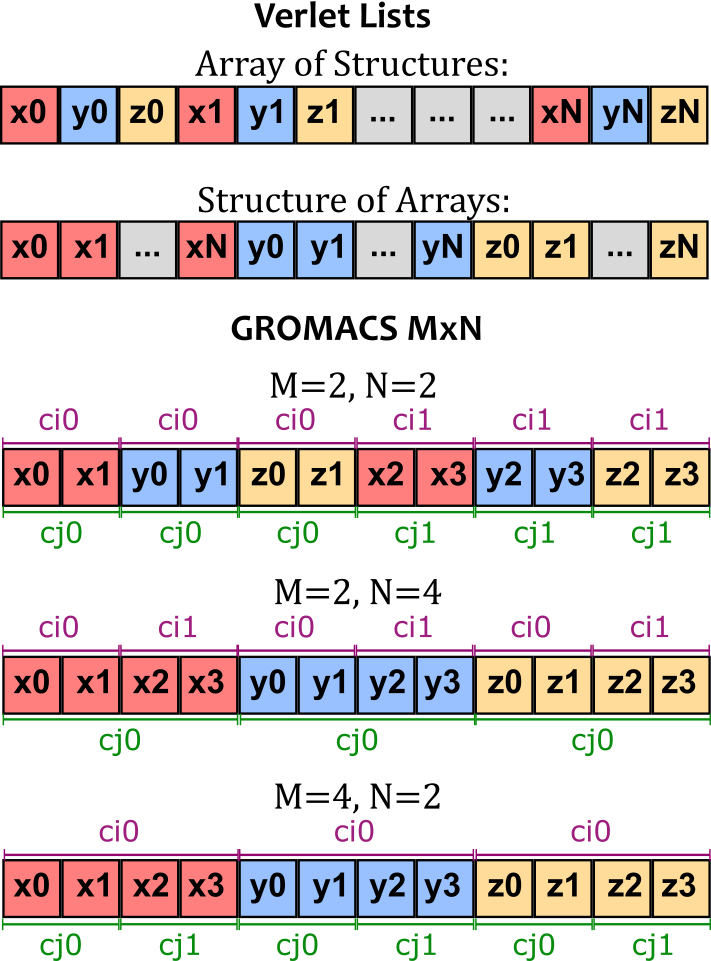}
	\caption{Data layouts for Verlet Lists and GROMACS MxN cases. Colors indicate properties in the same dimension. For Verlet Lists, note the spatial locality for one atom's coordinates in the \ac{AoS} case. For MxN, data is grouped by $i$-clusters or $j$-clusters (the one with higher size). References between data regions and which clusters they belong to are also displayed in purple (for $i$-clusters) and green (for $j$-clusters).}
    \label{fig:mdlayouts}
\end{figure}

The algorithm has two major drawbacks with respect to SIMD parallelism: (a) an irregular access pattern, since neighbor atoms are scattered across memory and (b) no reuse of neighbor data (positions) across atom iterations.
As a consequence, a gather operation is required to load the neighbor atoms' data into the vector registers. 
There are two strategies to gather data, namely \emph{hardware} or \emph{software} gathers.
Hardware gathers on x86-64 processors use a \emph{vgather} instruction
to perform the gathering at hardware level, where software gathers emulate the gather operation using separate instructions to load, shuffle and permute elements within vector registers.
Atom properties are typically stored in an \ac{AoS} layout, meaning that coordinates for one atom in different dimensions (X, Y and Z) are consecutively stored in memory.
This is the opposite layout of \ac{SoA} layout, where all atoms' coordinates in one dimension are contiguous in memory, with no spatial locality on coordinates from the same atom.
Figure~\ref{fig:mdlayouts} depicts how data is stored in memory for both layouts.

When available in the target processor, hardware gathers clearly use fewer instructions but cannot take advantage of spatial locality in the \ac{AoS} layout, thus requiring at least three data transfers instead of one when keeping elements aligned to the cache line size.
In other words, software gathers load the atom coordinates from one neighbor into one vector register with a single load instruction, thus requiring a single memory transfer (instead of three) when data is properly aligned to the cache line size.
Due to this trade-off between instruction execution and memory transfer, it is not straightforward to determine the best strategy, as it will depend on the target processor.

For all kernel variants available, MD-Bench contains their \emph{half neighbor lists} counterparts. These variants build the neighbor lists in only one direction and take advantage of Newton's Third Law to reuse the forces computed during pair interactions for the neighbor atom. When partial forces are computed, they are summed up into the force of the current atom and subtracted from the force of the neighbor atom (see Algorithm~\autoref{alg:forcehn}).
Despite the clear benefit of roughly halving the amount of pair interactions to compute, this strategy harms parallelism because it introduces race conditions for the neighbor atoms: Forces are stored back to memory in the innermost loop.
Also, gather and scatter operations are needed for the forces of the neighbor atoms, which not only affects the instruction execution cost but also increases memory traffic.

\subsubsection{MxN cluster algorithm}
\label{sec:gromacs}


MD-Bench is the first \ac{MD} proxy-app to introduce the MxN algorithm implemented in the GROMACS package to calculate short-range non-bonded atomic forces.
The algorithm groups atoms in two levels: (a) the $i$-clusters (with size $M$), which correspond to atoms in the outermost loop of the force calculation, and (b) the $j$-clusters (with size $N$), which correspond to neighbor atoms in the innermost loop used to calculate the partial forces for the i-clusters.
$M$ parameterizes the reusability of data as atoms in the same $i$-cluster contain the same pair lists, and $N$ specifies how many neighbor atoms are loaded and computed in one cluster pair interaction, which results in $M \times N$ atom pair interactions computed per SIMD iteration.

Figure \ref{fig:gromacs_mxn} depicts an example for $M=4, N=4$, note that the range used to search neighbors for the $i$-cluster is the union of the search ranges from each of its individual atoms in the Verlet list case.
If at least one atom in a $j$-cluster is within the cutoff region, the whole $j$-cluster is inserted into the neighbor list, which may incur in waste of computational resources.
Also, when there are not enough atoms to fill in a cluster, it is completed with dummy atoms placed in the infinity to fail the cutoff checking, which also increases redundant computation.
Therefore, the trade-off between extra computations and efficiency for SIMD parallelism must be assessed, and $M$ and $N$ need optimal adjustment to reach ideal performance.

Besides reusing loaded data from memory across atoms in the same $i$-cluster, another advantage of the algorithm is the elimination of gather operations for atoms in the $j$-clusters.
The atoms' data in the clustered format is stored in groups of $\max(M,N)$, constituting an \ac{AoSoA} layout (see Figure~\ref{fig:mdlayouts}), thus data for atoms in the same cluster can be loaded with a single load instruction.

\begin{figure}
    \centering
    \includegraphics[width=8.5cm]{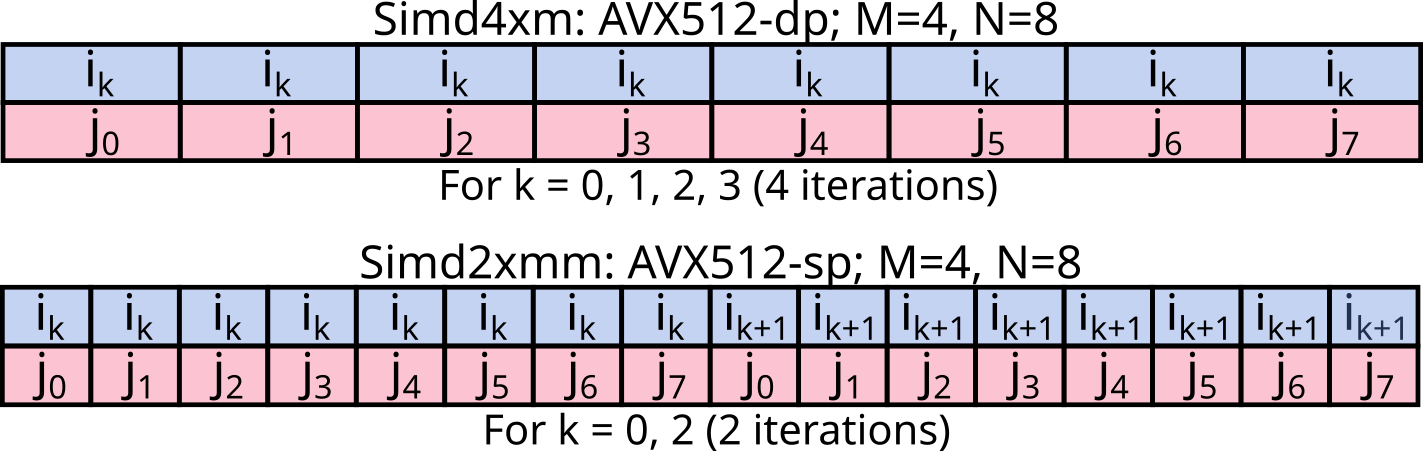}
    \caption{SIMD vector layout for Simd4xm and Simd2xmm pair interaction kernels on AVX512 machines. Simd4xm requires M SIMD iterations to compute all interactions (unroll factor = M) and Simd2xmm requires $\frac{M}{2}$ SIMD iterations since one AVX512 register is able to compute $N\times2$ interactions per time when using single-precision data types (unroll factor = $\frac{M}{2}$).}
    \label{fig:gromacskernels}
\end{figure}

To make use of efficient SIMD parallelism, tailored kernels with vector extension intrinsics are required.
Consider $W$ the amount of elements a SIMD register can hold, which depends on the vector width of the CPU in bytes and the floating-point precision used (for AVX512, $W=8$ for double-precision and $W=16$ for single-precision floating-point elements).
Available CPU kernels assume $M = 4$, and $N$ is set based on $W$.
The loop over the $i$-cluster is unrolled to compute every pair interaction, and the unroll factor depends on how many interactions between $i$-clusters and $j$-clusters can be computed using one SIMD register.
To demonstrate this, we describe the three kernel variants available for CPU targets (see Figure~\ref{fig:gromacskernels} for Simd4xm and Simd2xmm examples):

\begin{itemize}
  \item \textbf{Reference:} uses raw loops instead of intrinsics to iterate over atoms in clusters; available for validation purposes.
  \item \textbf{Simd4xm:} $M=4$, $N=W$; j-cluster elements occupy full SIMD register, thus each i-cluster data needs to be loaded into a separate SIMD register (unroll factor = 4); selected when using double-precision with AVX512
  \item \textbf{Simd2xmm:} $M=4$, $N=W/2$; j-cluster elements occupy half a SIMD vector, thus elements from two i-clusters can share the same SIMD register (unroll factor = 2); selected when using single-precision with AVX512
\end{itemize}

Note that in all cases $M$ is either $\frac{N}{2}$, $N$ or $2N$.
Since a bidirectional mapping between $i$-clusters and $j$-clusters is needed in many procedures of the GROMACS algorithm, keeping this relation between $M$ and $N$ makes the mapping much simpler and also avoids significant overhead.

\subsection{Benchmark test cases}\label{sec:testcases}
Short-range force kernels for \ac{LJ} and \ac{EAM} potentials are available in MD-Bench.
Setups can be provided by reading atom positions (and velocities in some cases) in a Protein Data Bank (PDB), Gromos87 or LAMMPS dump file.
The following test cases from material modeling and bio-sciences fields are available and used in this paper:
\begin{itemize}
  \item \textbf{Standard:} Standard copper FCC lattice case from material modeling used in miniMD; 131072 atoms, with an average of 76 neighbors per atom and 200 time-steps.
  \item \textbf{Melt:} Similar to Standard case, but the material melts over time; 32000 atoms, with an average of 76 neighbors per atom and 200 time-steps.
  \item \textbf{Argon:} Pure-argon simulation (bio-sciences); 1000 atoms, with an average of 64 neighbors per atom and 250000 time-steps.
\end{itemize}
Note that long-range potentials are not implemented in MD-Bench, hence test cases are limited to simulations without long-range effects such as electrostatic forces.
Since our goal is not to develop a production \ac{MD} application but rather analyze the performance from short-range potential kernels, this is not an issue as long as we can provide test cases for multiple fields and with different characteristics that are interesting from the performance point of view.

\subsection{Tools}\label{sec:tools}

Besides the transparent and simpler implementations of \ac{MD} kernels, MD-Bench also contains many tools to extract performance data and isolate distinct contributions during execution on a specific target hardware.
The following is a non-exhaustive list of these tools:
\begin{itemize}
  \item \textbf{Detailed statistics mode} To collect extended runtime information for the simulation, a detailed statistics mode can be enabled.
It introduces various statistics counters in the force kernels in order to collect relevant metrics such as cycles per pair interaction (processor frequency must be fixed), average cutoff conditions that fail/succeed and useful read data volumes.

  \item \textbf{Memory traces and gather-bench} The gather benchmark is a standalone benchmark code for (currently only) x86-64 CPUs that mimics the data movement (both operations and transfers) in \ac{MD} kernels in order to evaluate the ``cost of gather'' on different architectures.
It currently gathers data in the following patterns: (a) simple 1D arrays with fixed stride, to evaluate single gather instruction in the target CPU, (b) array of 3D vectors with fixed stride, to evaluate \ac{MD} gathers with regular data accesses and (c) array of 3D vectors using trace files from MD-Bench \emph{INDEX\_TRACER} option, to evaluate \ac{MD} gathers with irregular data accesses.
The benchmark exploits the memory hierarchy by adjusting the data volume to fit into a different cache level in successive executions, and yields a performance metric based on the number of cache lines touched.
There are options to determine the floating-point precision and to include padding to ensure alignment to the cache line size.

  \item \textbf{Stubbed force calculation}\label{sec:stubbed} To execute \ac{MD} kernels in steady state and understand their performance characteristics, we established a synthetic ``stubbed'' case where the number of neighbors per atom remain fixed and the data access pattern can be predicted.
Figure~\ref{fig:stub} depicts examples of data access patterns available for $N=4$ neighbors.
The patterns are listed as follows:
\begin{itemize}
    \item \textbf{sequential:} every atom has its $N$ subsequent atoms as its neighbors; neighbor lists vary across atoms and data access is contiguous in memory.
    \item \textbf{fixed:} every atom has the first $N$ atoms in the simulation as its neighbors; neighbor lists are the same for all atoms and data access is contiguous in memory.
    \item \textbf{random:} every atom has $N$ random atoms as its neighbors; neighbor lists vary across atoms and data access is scattered over memory.
\end{itemize}
Both irregular data accesses and variations in the innermost loop size can be addressed by the benchmark. It can therefore be used to isolate contributions from memory latency and control-flow divergence.

For GROMACS kernels, the patterns are analogous but clusters substitute atoms.
The amount of atoms in the simulation is specified by the number of $i$-clusters and the number of neighbors per $i$-cluster is specified by the number of $j$-clusters.
There is also an option to specify the number of atoms per $i$-cluster to evaluate the performance impact when clusters are not fulfilled (i.e., when they are partially filled with dummy atoms).
\end{itemize}

\section{Examples}\label{sec:showcases}

\subsection{Materials and methods}\label{sec:testbed}


\begin{table*}[htb]
  \centering
  \begin{tabular}{c|c|c|c}
    Micro-architecture & Cascade Lake & Ice Lake server & Zen3 \\
    \hline
    Model & Xeon Gold 6248 & Xeon Platinum 8360Y & EPYC 7543 \\
    Base frequency & 2.4\,\GHZ & 2.5\,\GHZ & 2.1\,\GHZ \\
    Cores per chip & 20 & 36 & 32 \\
    Latest SIMD extension& AVX512 & AVX512 & AVX2+FMA \\
    L1D capacity & $20 \times 32$\,\KiB & $36 \times 48$\,\KiB & $32 \times 32$\,\KiB \\
    L2 capacity & $20 \times 1024$\,\KiB & $36 \times 1.25$\,\MiB & $32 \times 512$\,\KiB \\
    L3 capacity & 27.5\,\MiB & 54\,\MiB & 256\,\MiB \\
    Memory configuration & 6 ch. DDR4-2933 & 8 ch. DDR4-3200 & 8 ch. DDR4-3200 \\
  \end{tabular}
  \caption{Specifications for testbed CPUs.}
  \label{tab:testbedcpus}
\end{table*}

\begin{table*}[htb]
  \centering
  \begin{tabular}{c|c|c|c}
    Kernel & Target & Compiler & Flags\\
    \hline
    lammps-avx512-dp & Cascade Lake & ICC 2021.6.0 & \verb+-Ofast -xCORE-AVX512 -qopt-zmm-usage=high+ \\
    lammps-avx2-dp & Cascade Lake & ICC 2021.6.0 & \verb+-Ofast -xCORE-AVX2+ \\
    gromacs-avx512-sp/-dp & Cascade Lake & ICX 2021.1 Beta & \verb+-Ofast -xCORE-AVX512 -qopt-zmm-usage=high+ \\
    lammps-avx2-dp & Zen3 & ICX 2021.1 Beta & \verb+-Ofast -xHost+ \\
    lammps-novec-dp & Ice Lake server & ICC 2021.6.0 & \verb+-Ofast -no-vec+ \\
    lammps-sse-dp & Ice Lake server & ICC 2021.6.0 & \verb+-Ofast -xCORE-SSE4.2+ \\
    lammps-avx2-dp & Ice Lake server & ICC 2021.6.0 & \verb+-Ofast -xCORE-AVX2+ \\
    lammps-avx512-dp & Ice Lake server & ICC 2021.6.0 & \verb+-Ofast -xCORE-AVX512 -qopt-zmm-usage=high+ \\
    gromacs-avx512-sp/-dp & Ice Lake server & ICX 2021.1 Beta & \verb+-Ofast -xCORE-AVX512 -qopt-zmm-usage=high+ \\
    gromacs-avx512-sp/-dp & Ice Lake server & ICX 2022.1.0 & \verb+-Ofast -xCORE-AVX512 -qopt-zmm-usage=high+ \\
    gromacs-avx512-sp/-dp & Ice Lake server & GCC 12.1.0 & \verb+-Ofast -mavx512f -mavx512vl -mavx512bw+ \\
    &&& \verb+-mavx512dq -mavx512cd -ffast-math+ \\
    &&& \verb+-funroll-loops+ \\
    gromacs-avx512-sp/-dp & Ice Lake server & CLANG 15.0.1 & \verb+-Ofast -march=icelake-server+ \\
    &&& \verb+-mprefer-vector-width=512+ \\    
  \end{tabular}
  \caption{Compiler and flag specifications for different kernels used in this paper.}
  \label{tab:testbedcompilers}
\end{table*}


In this work we conducted experiments on two Intel CPUs (with Cascade Lake and Ice Lake server micro-architectures) and on an AMD Milan (Zen3) CPU.
The CPUs and their specifications are listed in \autoref{tab:testbedcpus}.
\autoref{tab:testbedcompilers} lists the compilers and flags used for the different kernels.
The fastest versions were generated by either the Intel C++ Compiler Classic (ICC) 2021.6.0 20220226 or the Intel oneAPI DPC++/C++ Compiler (ICX) 2021.1 Beta 20200304.
For the sake of simplicity, we will refer to these as ICC and ICX, respectively.

In \autoref{sec:latency} we used static code analysis tools to obtain predictions that serve as optimistic lower limits.
The tools used are IACA v3.0-28-g1ba2cbb \cite{iaca}, OSACA v0.4.13 \cite{8641578}, llvm-mca v15.0.6 \cite{llvm-mca} and uiCA \cite{Abel22}.

Unless otherwise stated, all experiments were executed with fixed frequency (set to the base clock speed) and pinned to a single core.
We used LIKWID (V5.2) to fix the CPU frequency, pin tasks to specific cores, enable/disable prefetchers and make use of \ac{HPM} counters. Prefetcher settings on the AMD CPU are currently not supported by LIKWID; in this case we modified the BIOS settings of the benchmark machine.

\subsection{Floating-point precision analysis}\label{sec:fpanalysis}

The use of single-precision (SP) floating-point arithmetic for the force calculation is not unusual in \ac{MD} simulations.
For instance, GROMACS is compiled with mixed precision by default instead of using double precision (DP) for all variables and arrays.
In this scenario, the atom properties such as mass, position and velocity are stored and computed in SP, but some critical variables such as the virial (which is the sum of all forces in the system) are in DP.
This means that at least in some cases the computation of individual forces for each atom can be done entirely in SP, and only macroscopic properties and the results of reduction operations (such as the virial) need to be stored and computed in DP.
It is therefore important to understand the performance impact of using SP in such force kernels, and we focus on addressing these implications here.
We evaluate the use of SP for both LAMMPS and GROMACS algorithms and measure different resources in the target CPU (in this case Cascade Lake) with LIKWID\@.
\begin{table*}[htb]
    \centering
    \begin{tabular}{c|c|c|c|c}
        Variant & lammps-sp & lammps-dp & gromacs-sp & gromacs-dp \\
        \hline
        Runtime [\seconds] & 3.78 & 4.37 & 3 & 5.23 \\
        AVX512 performance [\GFS] & 16.6209 & 14.4449 & 52.4009 & 28.9085 \\
        CPI & 1.2189 & 0.8292 & 0.3778 & 0.4703 \\
        Total instructions $(\times 10^9)$ & $7.4284$ & $12.5921$ & $18.9560$ & $26.7498$ \\
        FP instructions $(\times 10^9)$ & $4.014$ & $8.155$ & $10.0616$ & $19.8296$ \\
        Arithmetic instructions ratio [\%] & 54.03 & 64.76 & 52.97 & 74.05 \\
        Vectorization ratio [\%] & 98.0310 & 99.0308 & 100 & 100 \\
        L2/L1 load data bw [\GBS] &	3.3335 & 4.6478 & 2.6965 & 2.7822 \\
        L2/L1 load data vol. [\GB] & 14.3856 & 20.8334 & 8.2727 & 14.9912 \\
        L3/L2 load data bw [\GBS] & 3.233 & 3.2462 & 1.2325 & 1.2310 \\
        L3/L2 load data vol. [\GB] & 12.6089 & 14.551 & 3.7813 & 6.633 \\
        Memory read data bw [\GBS] & 1.9953 & 1.9754 & 0.1587 & 0.2190 \\
        Memory read data vol. [\GB] & 7.5571 & 8.8299 & 0.4864 & 1.1955 \\
    \end{tabular}
    \caption{Measurements from LAMMPS and GROMACS force kernels using different floating-point precision on Cascade Lake (AVX512, Standard benchmark case).}
    \label{tab:results_fp_analysis}
\end{table*}

\begin{table*}[htb]
    \centering
    \begin{tabular}{c|c|c||c|c|c|cc}
        Kernel (stubbed) & Target & Meas. & IACA & uiCA & LLVM-MCA & OSACA & naive Frontend \\
        \hline
        lammps-avx512-dp & Cascade Lake & 4.10 & 3.86  & 2.96  & 1.62 & \textbf{2.35} & 1.78 \\
        lammps-avx2-dp   & Cascade Lake & 7.25 & 6.39  & 5.43  & 2.62 & 3.42 & \textbf{3.87} \\
        gromacs-avx512-dp& Cascade Lake & 1.62 & 1.53  & 1.53  & 1.06 & \textbf{1.37} & 0.96 \\
        gromacs-avx512-sp& Cascade Lake & 0.90 & 0.87  & 0.84  & 0.53 & \textbf{0.70} & 0.67 \\
        lammps-avx512-dp & Ice Lake server    & 3.26 & 3.86  & 3.58  & 1.62 & \textbf{2.75} & 1.78 \\
        lammps-avx2-dp 	 & Ice Lake server    & 5.97 & 6.39  & 4.36  & 2.62 & 3.20 & \textbf{3.87} \\
        gromacs-avx512-dp& Ice Lake server    & 1.51 & 1.53  & 2.37  & 1.06 & \textbf{1.37} & 0.96 \\
        gromacs-avx512-sp& Ice Lake server    & 0.83 & 0.87  & 1.18  & 0.53 & \textbf{0.70} & 0.67 \\
        lammps-avx2-dp 	 & Zen3         & 8.53 & ---   & ---   & 5.37 & 4.70 & \textbf{4.78} \\
    \end{tabular}
    \caption{Measurements from stubbed case and in-core predictions using several analysis tools of different force kernels and micro-architectures. All timings are reported in cycles per pair interaction. The bold number in the OSACA+naiveFrontend columns indicates the maximum of the two. Note that neither IACA and uiCA support non-Intel architectures and, thus, cannot provide data for Zen3.}
    \label{tab:incore-prediction}
\end{table*}

\autoref{tab:results_fp_analysis} shows the obtained measurements with the Standard case.
We observe that using SP for the LAMMPS algorithm leads to about 15\% lower runtime in comparison with DP\@.
Based on the data access pattern from the LAMMPS kernel, there is no significant memory data traffic reduction when using SP because the only elements we can assure to be contiguous in memory and be loaded in one cache line are the coordinates for the same atom, which are just three elements and this assumption already holds when using double-precision values.
In other words, the read cache lines are only partially used, and the fraction of useful data is not increased when employing SP.
This assumption is corroborated by the measured read data volume, which only decreases by about 15\% for the main memory and by about the same ratio for the L3 to L2 cache transfers, hence it does not get even close to the factor of two as one would naively expect.
The only data volume with a significant reduction is the volume from L2 to L1 caches (45\%), which means there is less pressure on the first-level caches when using SP.

With SP, the overall number of instructions goes down by 41\%, while the reduction is 51\% for floating-point arithmetic instructions.
The reciprocals $\frac{1}{x_{ij}}$ of distances between atoms need to be computed (see \autoref{eq:lennard_jones}), which requires either an approximate reciprocal or division instruction.
Approximate reciprocal instructions yield results with 14-bits accuracy, and they can achieve almost the same precision as the division instruction by computing iterations of the Newton-Raphson (NR) method~\cite{intelOptManual}.
For DP, two NR iterations are computed to achieve an accuracy of 52 bits, while for SP only one NR iteration is performed to achieve an accuracy of 23 bits; in practice, this means there is one less \verb+vfmadd213+ instruction for the SP case.
Since ICC generates approximate reciprocal instructions with NR iterations, this explains why the reduction in the arithmetic instructions for SP is more than the expected 50\%.
The vectorization ratio is close to 100\% in both cases, and the SP instructions in the innermost loop are equivalent to the ones in the DP kernel.
Note that the \verb+vgatherspd+ \footnote{\url{https://www.uops.info/html-instr/VGATHERDPS_ZMM_K_VSIB_ZMM.html}} instruction used in the SP kernel has a higher latency and lower throughput compared to the \verb+vgatherdpd+ \footnote{\url{https://www.uops.info/html-instr/VGATHERDPD_ZMM_K_VSIB_YMM.html}} employed in the DP kernel, which can be one of the reasons the performance improvement is still far from a factor of two and also explains the higher \ac{CPI}.
There should be also more masked-out arithmetic operations for the SP case since one SIMD iteration computes twice the amount of atom-pair interactions.



For the GROMACS algorithm, SP shows a significant performance improvement over DP\@.
The overall data volume from main memory is reduced by more than a factor of two (59\%), which is expected since the neighbor atoms' data is contiguous for atoms within the same cluster and can be loaded with fewer cache line transactions.
Besides that, both SP and DP cases are fully vectorized as expected due to the explicit use of SIMD intrinsics.
Note that these are different force kernels for GROMACS (Simd4xm for DP and Simd2xmm for SP); for the SP case, some extra instructions are still required to split a vector register into two parts, which is probably the reason why the observed SP speedup is less than two. This is also confirmed by the arithmetic instructions ratio, which is considerably higher for DP\@.

\subsection{Assembly code analysis}\label{sec:asm}

Here we show how to use MD-Bench to evaluate the optimizations performed by compilers at the instruction code level in order to pinpoint possible improvements in the compiler-generated code.

The ICC compiler generates the best code for the LAMMPS algorithm on Intel CPUs supporting both AVX2 and AVX512 instruction sets.
For AVX512, the presence of an \emph{epilogue variant} for the innermost loop that actually uses AVX512 instructions is one of the main reasons, other compilers generate scalar code to compute the remainder elements.
In more detail, ICC produces three variants of the innermost loop with no unrolling beyond SIMD. If $n_i$ is the number of neighbors of the current atom $i$ and $k$ is the number of interactions already computed for it, then $r = n_i - k$ is the number of remaining neighbors to be computed. With $n_\mathrm{SIMD}$ being the number of doubles per vector (here 8 for AVX512 and 4 for AVX2) and $n_t$ being a threshold (1200 for AVX512 and 600 for AVX2), these are the code variants:
\begin{itemize}
  \item $r < n_\mathrm{SIMD}$: epilogue or last iteration; SIMD registers are not completely filled and unused slots are filtered out using mask registers for AVX512; this variant is not SIMD-vectorized for AVX2.
  \item $r \in \left]n_\mathrm{SIMD}, n_t\right]$: intermediate number of iterations; masks or conditionals are not required to check if neighbors are valid; this variant is SIMD-vectorized with the latest SIMD extension available for both AVX2 and AVX512.
  \item $r \geq n_t$: although generated by the compiler, this variant is not used and can therefore be ignored in this paper.
\end{itemize}

When compiling for AVX512 targets, ICX generates remainder loops with SSE instructions, so they are not SIMD-vectorized.
It also emits a \verb+vdivpd+ instruction instead of a \verb+vrcp14pd+, which has higher latency, lower throughput and more \muops.
AOCC, GCC and CLANG also do not generate AVX512 instructions to compute all pair interactions, which is also observed when measuring the number of retired scalar, SSE, AVX and AVX512 instructions.

Most compilers generate the \verb+vgather+ instructions to perform gathering at hardware level for the LAMMPS algorithm.
The only exception is the ICX-generated code for Zen3 (AVX2), which uses \verb+vmovhpd+ and \verb+vmovsd+ instructions to load every coordinate individually ($3$ coordinates $\times$ $4$ neighbors, hence $12$ instructions in total) and then employs \verb+vextracti128+ and \verb+vinsertf128+ instructions to build up the 32-byte SIMD registers.
The generated code does not take advantage of the spatial locality of data to build up the registers, which can be inferred from the presence of one \verb+vmov+ instruction with a source memory operand for each coordinate to be loaded.
The ICC compiler is able to generate the \verb+vgather+ instructions with the \verb+-march=core-avx2+ flag when compiling for Zen3, but this code is slower than the variant just described.
Also, ICX was the best choice for Zen3 and GROMACS cases.


\definecolor{cssgreen}{rgb}{0.0, 0.5, 0.0}
\lstset{
   resetmargins=true,
   xleftmargin=0cm,
   xrightmargin=0cm,
   escapechar=!,
   basicstyle=\tiny,
   commentstyle=\ttfamily\tiny\color{cssgreen}
}

\begin{table}
  \centering
  \begin{tabular}{c|c|c|c}
    Kernel & T ($\times10^{-1}$) & P ($\times10^{-1}$) & Time (\seconds) \\
    \hline
    lammps-dp+NR & $7.961495$ & $6.721043$ & 4.37 \\
    lammps-sp+NR & $7.961477$ & $6.721027$ & 3.78 \\
    lammps-dp & $7.961635$ & $6.721161$ & 3.97 \\
    lammps-sp & $7.961648$ & $6.721172$ & 3.62 \\
    gromacs-dp & $7.961956$ & $6.721432$ & 5.26 \\
    gromacs-sp & $7.961976$ & $6.721449$ & 2.99 \\
  \end{tabular}
  \caption{Temperature (T), pressure (P) and force runtime measurements with and without Newton Raphson (NR) instructions on AVX512 kernels on Cascade Lake. Quantities are dimensionless and reflect \emph{lj} style from LAMMPS.}
  \label{tab:ccmp}
\end{table}

As discussed in \autoref{sec:fpanalysis}, instructions to calculate a few iterations of the Newton-Raphson (NR) method are generated after computing the approximate reciprocal; these are not present in kernels with explicit SIMD intrinsics such as the MxN kernels.
Their usefulness is arguable because other factors can change the results at this precision, like the order for partial forces calculation that varies significantly across different optimization and parallelization strategies.
\autoref{tab:ccmp} shows runtime, temperature and pressure for the LAMMPS Verlet Lists algorithm with and without the NR instructions, as well as for the GROMACS MxN algorithm.
Without NR, we can perceive a performance improvement of about 11\% and 4\% in the force calculation runtime for the LAMMPS algorithm when using double and single precision, respectively.
In terms of accuracy, not only are the differences for temperature and pressure small (in the order of $2\times 10^{-5}$), but even smaller than when comparing to the GROMACS cases.
In summary, it may be beneficial to accept a minor loss in precision in return for better performance when the variation in the results is negligible.

\subsection{Memory latency and control flow divergence contributions} \label{sec:latency}

\begin{figure*}[tb]
    \centering
    \subfigure[LAMMPS AVX512 double-precision]{\label{fig:latency_lammps_avx512_dp_cascadelake}\includegraphics[width=0.48\textwidth]{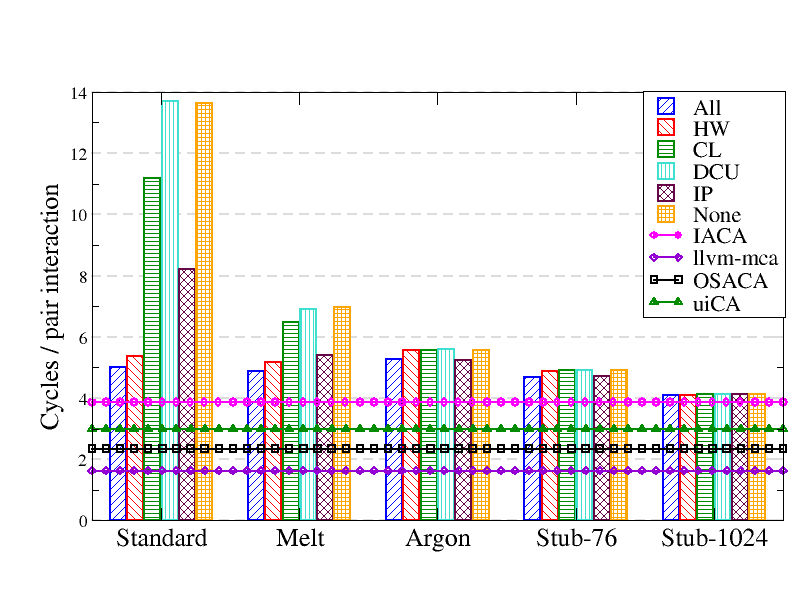}}
    \subfigure[LAMMPS AVX2 double-precision]{\label{fig:latency_lammps_avx2_dp_cascadelake}\includegraphics[width=0.48\textwidth]{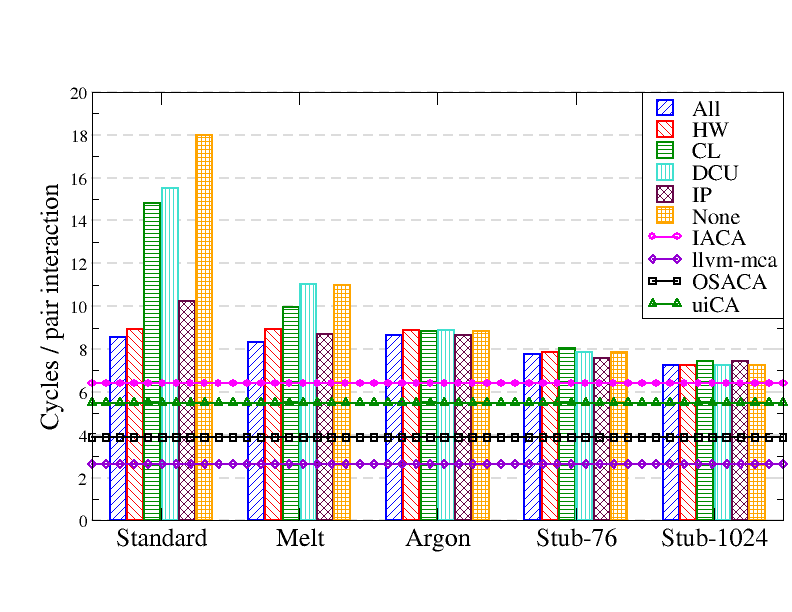}}
    \vskip\baselineskip
    \subfigure[GROMACS AVX512 double-precision]{\label{fig:latency_gromacs_avx512_dp_cascadelake}\includegraphics[width=0.48\textwidth]{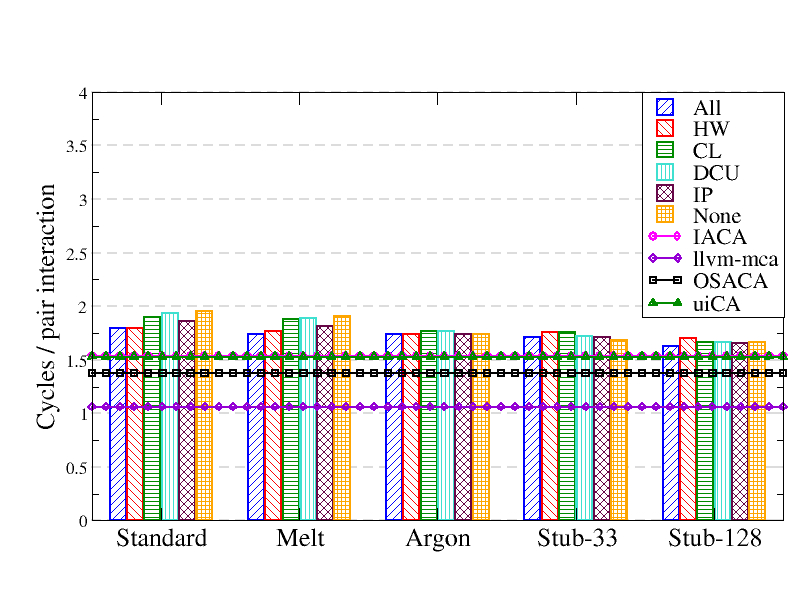}}
    \subfigure[GROMACS AVX512 single-precision]{\label{fig:latency_gromacs_avx512_sp_cascadelake}\includegraphics[width=0.48\textwidth]{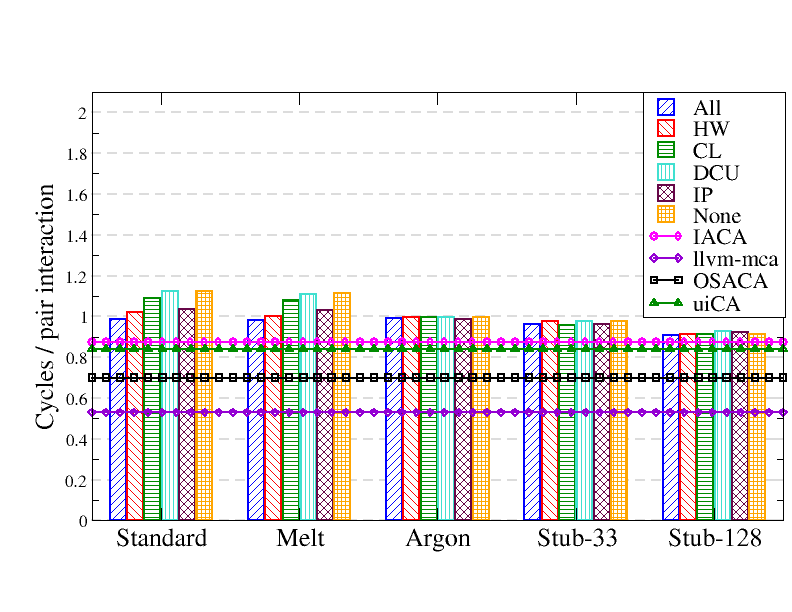}}
    \caption{Cycles per pair interactions measured for different test cases, with distinct prefetcher settings for the mentioned force kernels on Cascade Lake. Predictions from IACA (Skylake-X), llvm-mca, OSACA and uiCA are displayed in the y axis.}
    \label{fig:latencies_cascadelake}
\end{figure*}

\begin{figure*}[tb]
    \centering
    \subfigure[LAMMPS AVX512 double-precision]{\label{fig:latency_lammps_avx512_dp_icelake}\includegraphics[width=0.48\textwidth]{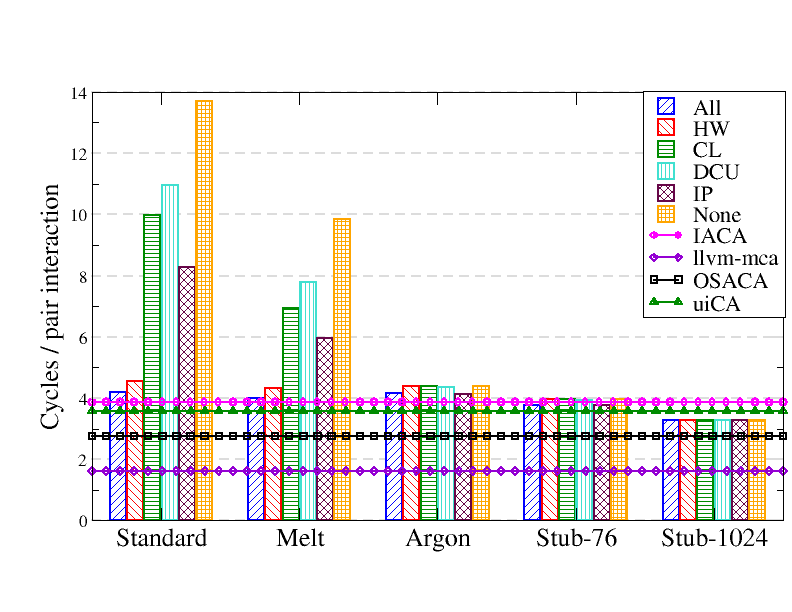}}
    \subfigure[LAMMPS AVX2 double-precision]{\label{fig:latency_lammps_avx2_dp_icelake}\includegraphics[width=0.48\textwidth]{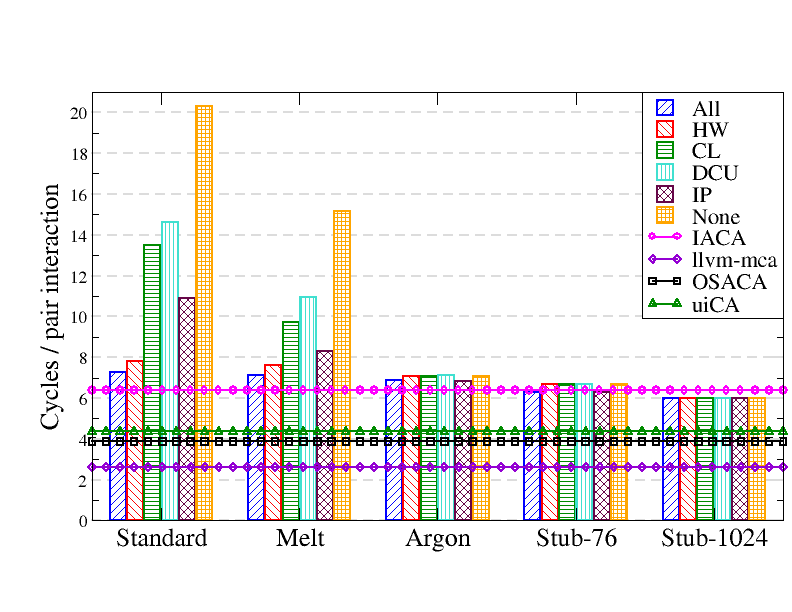}}
    \vskip\baselineskip
    \subfigure[GROMACS AVX512 double-precision]{\label{fig:latency_gromacs_avx512_dp_icelake}\includegraphics[width=0.48\textwidth]{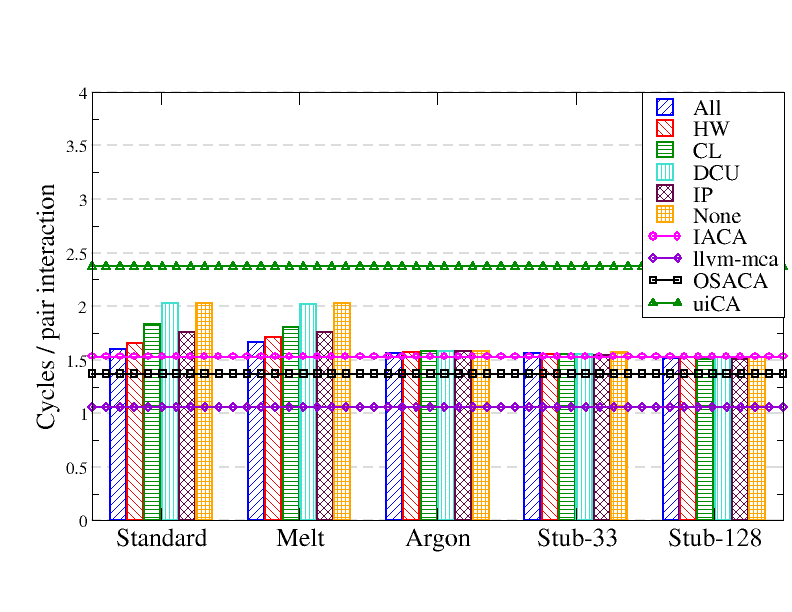}}
    \subfigure[GROMACS AVX512 single-precision]{\label{fig:latency_gromacs_avx512_sp_icelake}\includegraphics[width=0.48\textwidth]{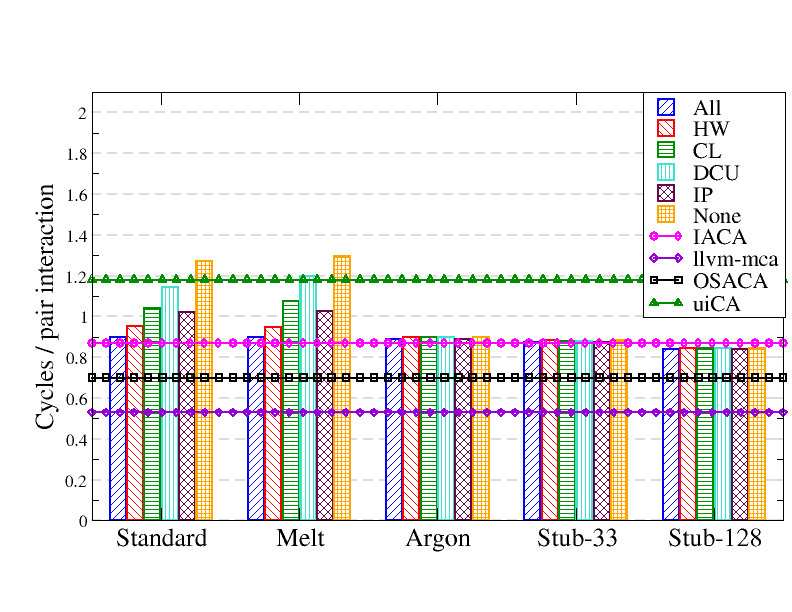}}
    \caption{Cycles per pair interactions measured for different test cases, with distinct prefetcher settings for the mentioned force kernels on Ice Lake server. Predictions from IACA (Skylake-X), llvm-mca, OSACA and uiCA are displayed in the y axis.}
    \label{fig:latencies_icelake}
\end{figure*}

\begin{figure}[tb]
    \centering
    \includegraphics[width=9cm]{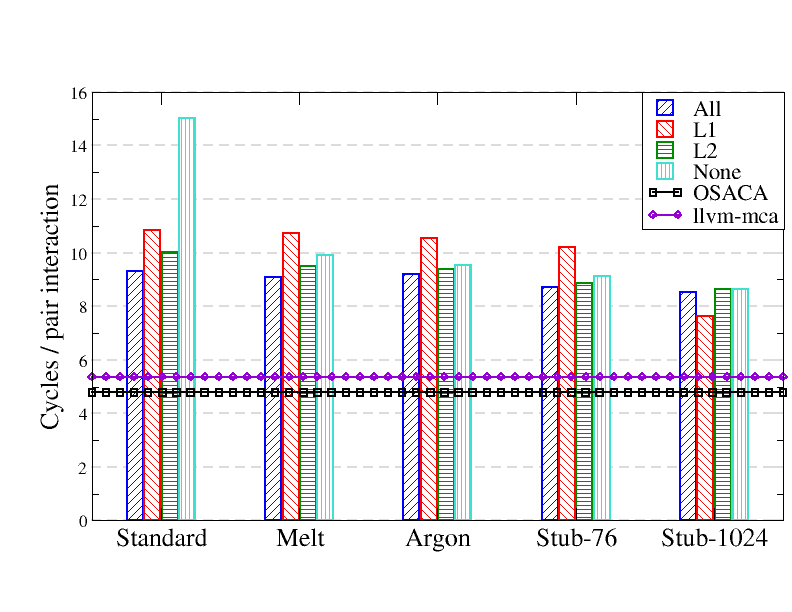}
    \caption{Cycles per pair interactions measured for different test cases, with distinct prefetcher settings for LAMMPS AVX2 double-precision kernel on Zen3. Predictions from IACA, llvm-mca, OSACA and uiCA are displayed in the y axis.}
    \label{fig:latency_lammps_avx2_dp_milan}
\end{figure}

An important assumption for performance engineering of streaming kernels is that latencies can be neglected due to a regular data access pattern, which cache prefetchers can handle very well.
At first glance, \ac{MD} simulations are expected to have significant latency impact due to their memory access characteristics, and MD-Bench can assist in measuring this impact via its stubbed sequential case.
Nonetheless, when comparing with our Standard, Melt and Argon cases, we observed that this effect is minor.
Besides latency, control flow divergence contributions are assessed by providing a stubbed version with a significantly higher number of neighbors per atom (1024 neighbor atoms for LAMMPS, 128 neighbor $j$-clusters for GROMACS), thus increasing the number of iterations on the innermost loop ($j$) before advancing to the outermost ($i$) loop's instructions.

Furthermore, we also compare measurements from the stubbed executions on Cascade Lake, Ice Lake and Zen3 with kernel throughput predictions under ideal conditions from IACA (configured for Skylake-X due to the lack of a newer option), OSACA, llvm-mca and uiCA static code analyzers, which are used as optimistic lower limits.
For experiments on Zen3, we only display OSACA and llvm-mca predictions since there is no support for AMD CPUs in IACA and uiCA.

Additionally, we extended the OSACA prediction, which only considers the backend, by a naive frontend analysis.
For this, we estimate an optimistic runtime by dividing the number of \muops\ (reported by llvm-mca) by the number of \muops\ the architecture can dispatch or retire per cycle.
This is 4 \muops\ per cycle for Cascade Lake and Ice Lake server~(bound by retirement throughput)~\cite{intelOptManual}, as well as for Zen3~(bound by dispatch throughput)~\cite{amdOptManual}. 
The overall runtime prediction for OSACA is then the maximum of the frontend and backend estimates.
Table~\ref{tab:incore-prediction} lists the measurements of the investigated stubbed versions and their predictions of the mentioned tools in cycles per pair interaction.
While the performance of the AVX512 variants is limited by the backend, the AVX2 variants suffer from additional instructions and are bound by the decoder throughput.
In the Ice Lake server case, IACA slightly overpredicts the actual measurements which contradicts the nature of a light-speed estimate, but since IACA predicts the runtime for the Skylake-X micro-architecture instead of Ice Lake server, i.e., an older CPU model, it can be assumed this is due to architectural differences.
A similar issue occurs with uiCA predictions for the Ice Lake server, the reason is uiCA only supports the Ice Lake client architecture which can execute only 1 FMA instruction per cycle when Ice Lake server can execute 2 FMA instructions per cycle.

Figures~\ref{fig:latencies_cascadelake} and \ref{fig:latencies_icelake} depict the cycle measurements on Cascade~Lake and Ice~Lake, respectively, for the different benchmark cases.
For the stubbed cases in all scenarios, the impact for disabling prefetchers is negligible as expected, the most significant increase of about 4\% is the LAMMPS AVX512 DP kernel with the Stub-128 benchmark, while other cases present a variation smaller than 1\%.
This suggests the stubbed case seems to succeed on suppressing the memory latency impact.

Figures~\ref{fig:latency_lammps_avx512_dp_cascadelake} and \ref{fig:latency_lammps_avx512_dp_icelake} show the measurements for LAMMPS AVX512 DP kernels.
With more irregular memory accesses, the average number of cycles grows by 7\% in the Standard case, 4\% in the Melt case and 12\% in the Argon case with all prefetchers enabled (blue bars).
Although these results (especially for the Argon case) display more significant impact of the memory latency than our previous work, it is still a minor contribution compared to the overall cycles spent.
When evaluating the stubbed versions with different amount of neighbors, the number of cycles decreases by about 12\% with 1024 neighbors per atom, hence control flow divergence is more significant for all cases except Argon, in which both contributions are roughly the same.
The prediction from IACA matches 94\% of the best execution, which points out our stubbed case seems close to optimal execution.
IACA reports stalled backend allocation in the CPU due to frontend bubbles, which matches our observations with the OSACA+naiveFrontend estimation. Note that in this case we assume no bubbles at all; the overall time needed for instruction decoding is therefore most likely even higher than what our model assumes.

Figures~\ref{fig:latency_lammps_avx2_dp_cascadelake} and \ref{fig:latency_lammps_avx2_dp_icelake} depict the measurements for LAMMPS AVX2 DP kernels.
The observed behavior is similar to AVX512, but with less impact from control flow divergence: about 6\% less cycles in relation to the overall, which is expected as the smaller vector width causes the number of iterations of the innermost loop to roughly double.
For the worse latency case (Argon), 10\% more cycles are needed compared to the stubbed case with 76 neighbors, hence latency impact is not significantly different from the AVX512 measurements.
The best prediction is also from IACA with 88\% fraction of the measurements, and a frontend model is made necessary to have good predictions for AVX2 kernels since OSACA prediction is smaller than the naive frontend estimation for these cases.

Figures~\ref{fig:latency_gromacs_avx512_dp_cascadelake}, \ref{fig:latency_gromacs_avx512_sp_cascadelake}, \ref{fig:latency_gromacs_avx512_dp_icelake} and \ref{fig:latency_gromacs_avx512_sp_icelake} depict the measurements from GROMACS AVX512 kernels for DP and SP.
In contrast to LAMMPS cases, the test-case with higher latency impact is Standard, with 4\% more cycles for DP and 2\% more cycles for DP in relation to the Stub-33 measurements.
Therefore, latency is even minor for GROMACS kernels which one could hypothesize based on the most efficient data access pattern.
Control flow divergence also represents a negligible contribution of 4\% more cycles for DP and 5\% more cycles for SP.
Predictions are more accurate compared to LAMMPS with IACA matching 94\% and 96\% fractions of the best measurement for DP and SP, respectively.

When ignoring stubbed cases, the most effective prefetcher for all kernels is the hardware prefetcher (HW), but the L1 IP prefetcher is also very successful on GROMACS kernels, and this is expected since it prefetches based on sequential load history and LAMMPS does not provide sequential accesses.
What can also be observed is that the hardware prefetcher substantially improves the results for the LAMMPS algorithm, which makes evident that there is a major concern for hiding latency on such kernels otherwise they can execute with 2-3$\times$ worse performance, while for GROMACS kernels such improvements are not so meaningful.

Figure~\ref{fig:latency_lammps_avx2_dp_milan} depicts the results for the LAMMPS AVX2 DP kernel on Zen3.
Besides the experiments with all and no prefetchers enabled, we split the available prefetchers into groups corresponding to the memory hierarchy level they operate on.
For the L1 group, we use the L1 stream, stride and region prefetchers, while for the L2 group only the L2 stream and up-down prefetchers are turned on.
It is noticeable that the L1 prefetchers are not so helpful when there are not many iterations on the innermost loop, even with stubbed measurements.
The reason is probably that these prefetchers work well with sequential data in the innermost loop (which only happens in the stubbed cases), but when jumping into the outermost loop the sequential pattern is interrupted by strided access from the next atom.
The L2 prefetchers have the opposite effect: they substantially improve the performance for the irregular accesses (especially for Standard), but do not show a consistent improvement when increasing the amount of neighbors.
On the contrary, the best performance is achieved with the Stub-1024 version with only the L1 prefetchers enabled, meaning that in this scenario L2 prefetchers actually harm the performance. This can happen due to cache pollution and extra data transfers from prefetched data that is not used.
Predictions are less accurate for AMD than for Intel. The closest prediction is from llvm-mca with 63\% of the measured cycles, which can be due to limitations in the tools or lack of a steady-state from the application in Zen3.

\subsection{Compiler code quality study I: HN versus FN algorithms using different SIMD instruction sets with the ICC compiler} \label{sec:codequality1}
\begin{figure*}[tb]
    \centering
    \subfigure[Runtime]{\label{fig:study_runtime}\includegraphics[width=0.32\textwidth]{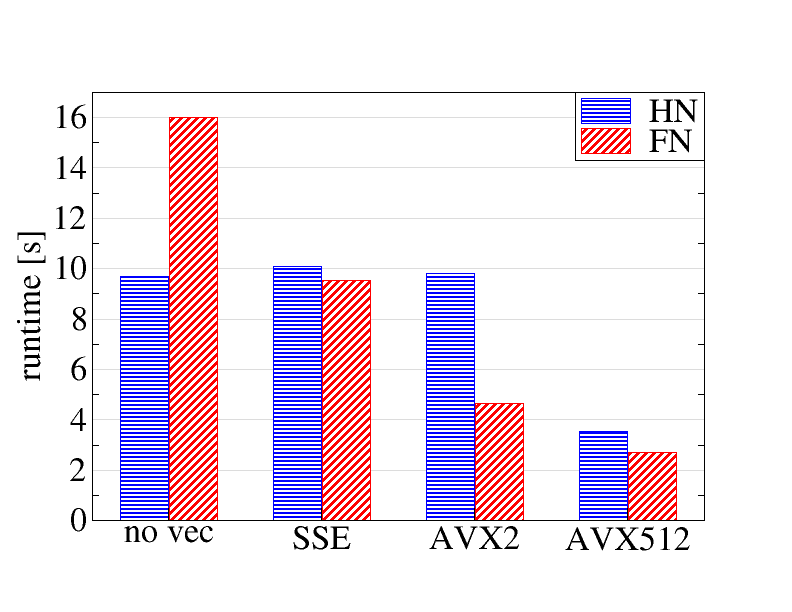}}
    \subfigure[HN Performance Profile]{\label{fig:study_hn_profile}\includegraphics[width=0.32\textwidth]{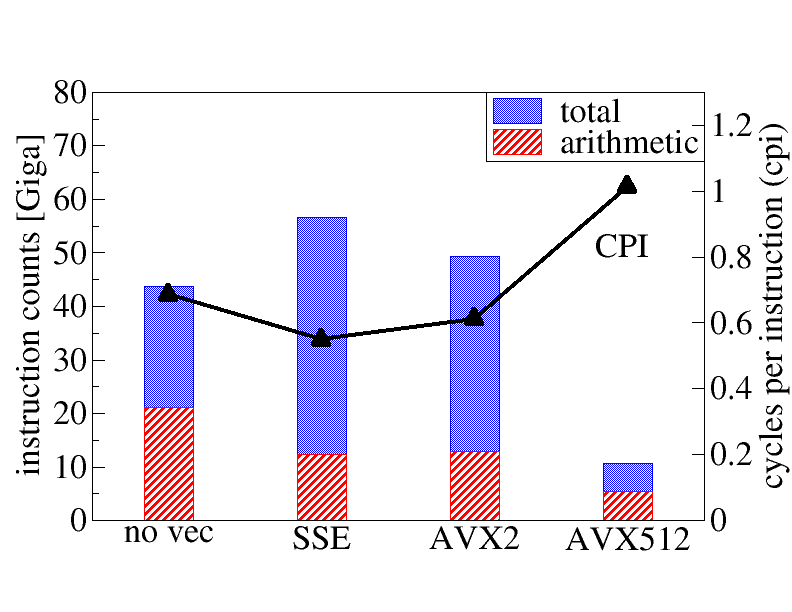}}
    \subfigure[FN Performance Profile]{\label{fig:study_fn_profile}\includegraphics[width=0.32\textwidth]{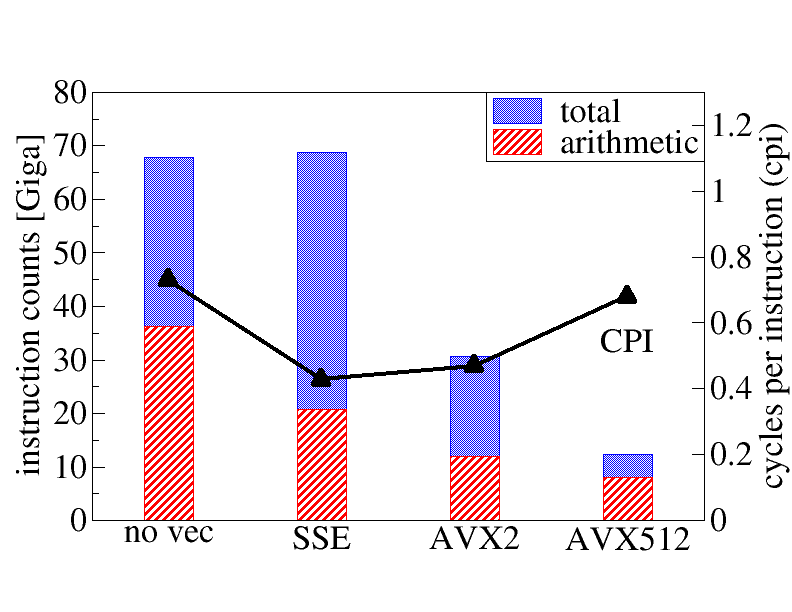}}
    \caption{Runtime (subfigure (a)) and HPM counter profiling results (half neighbor-list HN subfigure (b) and full neighbor-list FN subfigure (c)) for the Lennard-Jones copper lattice testcase. Results are shown using compiler flags to enforce no SIMD vectorization, SSE (16-byte), AVX2 (32-byte) and AVX512 (64-byte) SIMD vectorization. In subfigures (b) and (c) the stacked bars show total and arithmetic instruction counts on the left $y$-axis, the black lines denote cycles per instruction (CPI) on the right $y$-axis.}
\end{figure*}
In this analysis we benchmark and profile the Verlet List algorithm with \ac{HN} and \ac{FN} to see how well these are suited for vectorization and to explain the observed runtimes in more detail.
An \ac{AoS} data layout was used with double-precision floating-point arithmetic.
This study was performed on the Intel Ice Lake server node using the Intel compiler (ICC) 2021.6.0.
The force field kernel was compiled for several target \ac{SIMD} instruction sets and without vectorization using the \verb+-no-vec+ option.
The compiler requires a \verb+#pragma omp simd+ directive to vectorize the \ac{HN} variant.
The binaries were benchmarked with Turbo mode enabled, all executions were performed in the same cluster node with the same frequency, which was confirmed via HPM measurements.

Figure~\autoref{fig:study_runtime} shows the runtime for the Standard test case.
As expected, without vectorization the \ac{HN} variant is faster by almost a factor of two.
When using wider \ac{SIMD} units, \ac{FN} shows almost linear speed-up and is faster than \ac{HN} for all \ac{SIMD} widths.
\ac{HN} gets slower for SSE, stagnates for AVX2, and then improves considerably with AVX512 but still being 23\% slower than \ac{FN}.
For instruction throughput bound codes, the best variant is usually the one with  least instructions combined with optimal pipelined and superscalar execution, improving \ac{ILP}.
Both aspects can be directly measured using \ac{HPM} counters.
Figure~\autoref{fig:study_hn_profile} and Figure~\autoref{fig:study_fn_profile} show instruction counts and \ac{CPI} measurements for all \ac{HN} and \ac{FN} variants.
For \ac{HN} with SSE it can be seen that the arithmetic instruction count is almost half due to using the SSE (16 byte wide) registers.
Still, the compiler does not manage to reduce the overall instruction count.
29.6\% more instructions are required to get the operands into the \ac{SIMD} registers.
The additional instruction work is partially compensated by an improved \ac{CPI} resulting in only 3.7\% worse runtime.
An explanation for this improved \ac{CPI} is that the register/register SIMD instructions on Intel processors are executed on different scheduler ports than the arithmetic instructions and therefore can be executed out of order.
Although the compiler refused to employ full-width (32-byte) arithmetic \ac{SIMD} instructions in the AVX2 variant, the instruction count was still decreased and the runtime slightly improved.
The enhanced capabilities of the AVX512 instruction set extension enables the compiler to generate a version with just 25\% of the instruction count of the no-vec variant.
The runtime advantage is smaller because this instruction mix is executed with a significantly worse \ac{CPI} of $1.01$.
It is still impressive that a code that was impossible to vectorize efficiently with the previous \ac{SIMD} instruction set extensions now shows an instruction count reduction of a factor of four (out of the optimal eight).

For FN the compiler manages to reduce the arithmetic instruction count with every wider \ac{SIMD} unit.
The overall instruction count increased slightly for SSE, but then is just 45\% for AVX2 and 18\% for AVX512 compared to the no-vec variant.
This underscores that the FN version is very well suited for SIMD vectorization.

\subsection{Compiler code quality study II: SP versus DP with AVX512 using different compilers} \label{sec:codequality2}
\begin{figure*}[tb]
    \centering
    \subfigure[Runtime]{\label{fig:study_runtime_2}\includegraphics[width=0.32\textwidth]{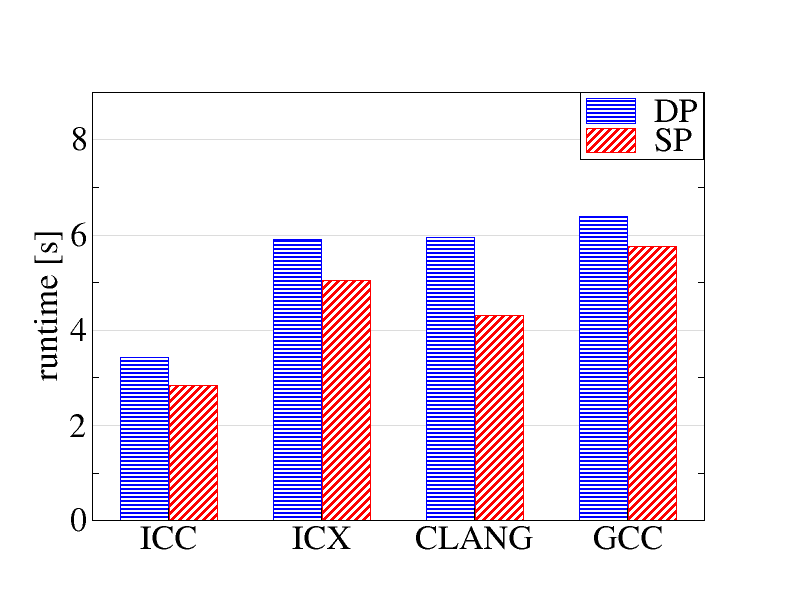}}
    \subfigure[DP Performance Profile]{\label{fig:study_dp_profile}\includegraphics[width=0.32\textwidth]{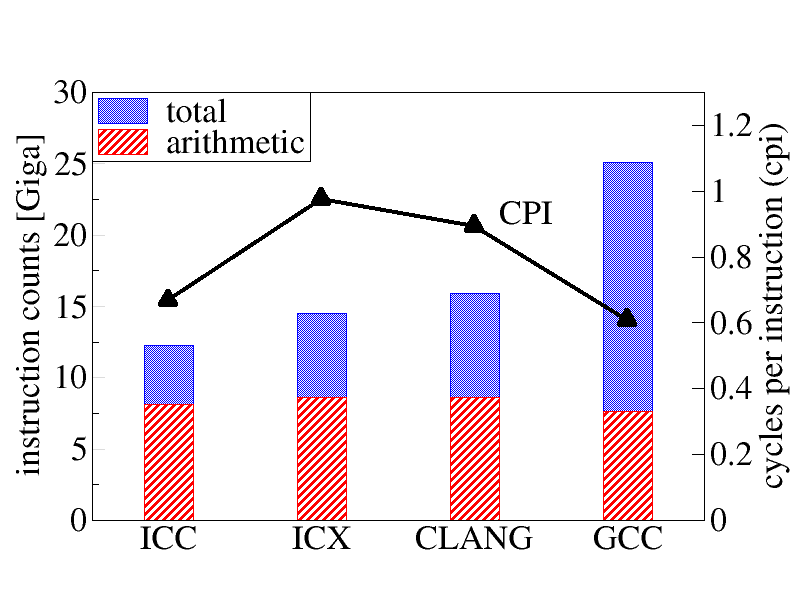}}
    \subfigure[SP Performance Profile]{\label{fig:study_sp_profile}\includegraphics[width=0.32\textwidth]{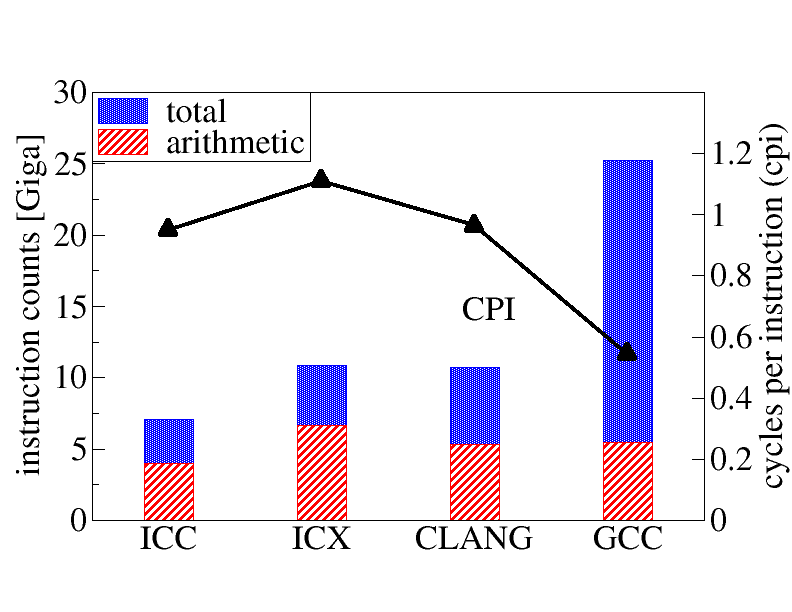}}
    \caption{Runtime (subfigure (a)) and HPM counter profiling results in double precision (DP) and single precision (SP) (subfigure (b) and  subfigure (c)) for the Lennard-Jones copper lattice testcase.
    Results are shown using using ICC 2021.6.0, ICX 2022.1.0, CLANG 15, and GCC 12.1.0 compilers with flags to enforce AVX512 (64-byte) SIMD vectorization.
    In subfigures (b) and (c), the stacked bars show total and arithmetic instruction counts on the left $y$-axis, the black lines denote cycles per instruction (CPI) on the right $y$-axis.}
\end{figure*}
\begin{figure*}[tb]
    \centering
    \subfigure[DP Performance Profile]{\label{fig:arith_dp_profile}\includegraphics[width=0.32\textwidth]{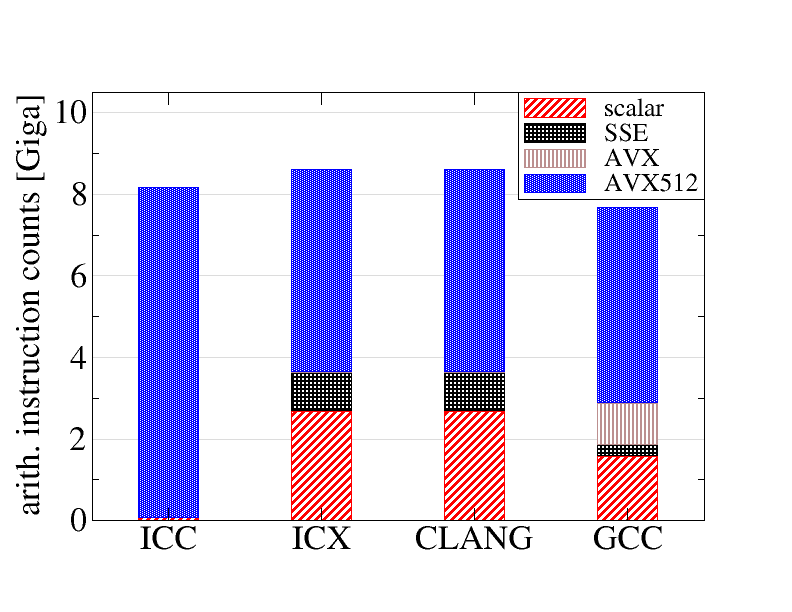}}
    \subfigure[SP Performance Profile]{\label{fig:arith_sp_profile}\includegraphics[width=0.32\textwidth]{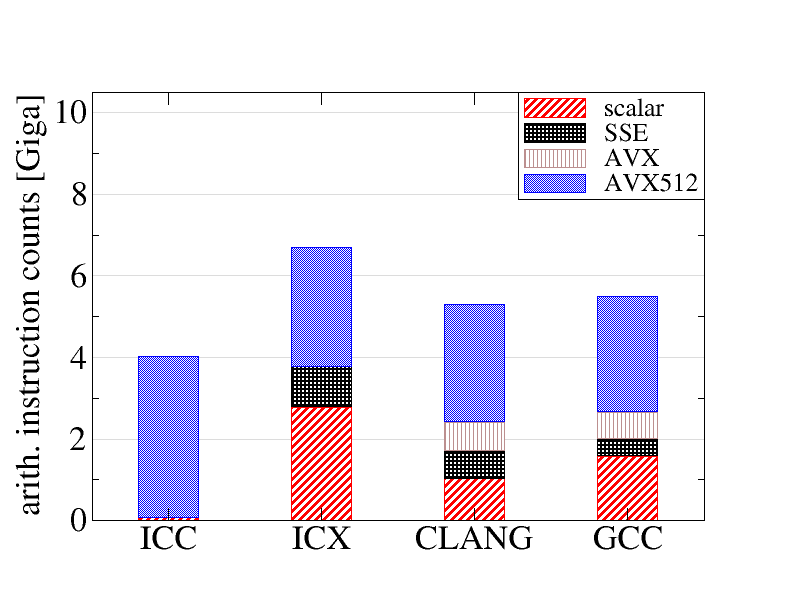}}
    \caption{HPM counter profiling results showing the arithmetic instructions decomposition for the Lennard-Jones copper lattice testcase (double precision (DP) in subfigure (a) and single precision (SP) in subfigure (b)).
    The stacked bars show the counts of scalar, SSE, AVX and AVX512 arithmetic instructions for the compilers ICC 2021.6.0, ICX 2022.1.0, CLANG 15, and GCC 12.1.0 using compiler flags to enforce AVX512 (64-byte) SIMD vectorization.
    }
\end{figure*}
In this analysis, the Verlet List algorithm with \ac{FN} and \ac{AoS} data layout is benchmarked using DP and SP versions and profiled to study how well different compilers can vectorize these cases using compiler flags to enforce AVX512 SIMD vectorization.
This was done on the Intel Ice Lake server node using compilers Intel ICC, Intel oneAPI ICX, LLVM CLANG, and GCC, with the corresponding compiler flags to enable AVX512 SIMD vectorization as listed in Table~\ref{tab:testbedcompilers}.
The clock frequency was fixed to 2.4\,GHz.
The Intel ICC compiler is known for its competitive auto-vectorizer.
Intel announced to deprecate their proprietary compiler in favor of an open-source LLVM-based compiler named ICX.
For comparison, this study also includes recent standard LLVM CLANG and GNU GCC compilers.
Unfortunately, it is not known which CLANG version the current Intel ICX compiler is based on, although it is expected to be older than CLANG 15.
The Verlet List algorithm is difficult to SIMD vectorize due to indirect accesses, an if condition in the innermost loop and the \ac{AoS} data layout used.
Advanced capabilities offered by the AVX512 instruction set extension such as gather instructions together with instruction masking and 32 floating-point registers allow to vectorize this algorithm easier than it was the case with previous SIMD instruction sets.
Figure~\autoref{fig:study_runtime_2} shows the runtimes for double-precision (DP) and single-precision (SP) using different compilers.
Assuming perfect SIMD vectorization, the SP case is expected to be twice as fast because the work can be executed by only half the number of instructions.
Additionally, the SP case also consumes only half of the data volume compared to DP.
However, the SP case is only 11\% faster for GCC, 17\% for ICX, 21\% for ICC, and 38\% for CLANG.
As can be seen in Figures~\autoref{fig:study_dp_profile} and \autoref{fig:study_sp_profile}, all compilers apart from GCC manage to reduce the overall instruction count, and ICC is even close to the optimal factor of two.
There are two possible explanations for the small speed-up between DP and SP:
First, while the SP gather instruction does twice the work it also requires significantly longer runtime compared to its DP counterpart (as discussed in \autoref{sec:fpanalysis}).
Second, in the SP case more cache line loads are triggered where not all the data can be used before eviction, causing a larger runtime contribution from data transfers than in the DP case (\autoref{tab:results_fp_analysis} shows the memory volume).
An indicator for both effects is the rise in CPI for the ICC compiler from $0.66$ in DP to $0.94$ in the SP case.

ICC produced by far the fastest code with a runtime of 3.43\,s for DP compared to 5.9\,s for ICX, 5.96\,s for CLANG, and 6.4\,s for GCC.
The situation is similar for SP with a runtime of 2.83\,s for ICC compared to 4.32\,s for CLANG, 5.04\,s for ICX, and 5.75\,s for GCC.
The ICC compiler is able to provide a very high quality AVX512 vectorization indicated by the almost 100\% vectorization ratio for arithmetic instructions and a high fraction of arithmetic versus total instructions of 66\% (DP) and 56\% (SP), which is an indicator of a low-overhead vectorization.
The two LLVM-based compilers ICX and CLANG show very similar performance for DP, with CLANG requiring slightly more total instructions compared to ICX.
The fact that the arithmetic instruction decomposition (see Figure~\autoref{fig:arith_dp_profile}) is almost identical with a significant scalar and SSE instruction fraction indicates that both compilers use the same vectorization backend.
GCC showed the highest total instruction count.
On the other hand, GCC required the fewest arithmetic instructions of all compilers, if with a small margin.
It also is the only compiler with a significant use of AVX arithmetic instructions in the DP case.
A detailed assembly code analysis (as done in \autoref{sec:asm}) would be required to understand the vectorization strategies the compilers employed.
For SP, CLANG chooses a different vectorization strategy than ICX with an arithmetic instruction mix closer to GCC (also using AVX instructions).
As a result, the CLANG compiler for SP produces 17\% faster code than ICX requiring fewer arithmetic instructions.
An explanation could be that the LLVM vectorizer was improved in LLVM 15 compared to the version ICX is based on.
CLANG shows a lower CPI compared to ICX in both cases.
GCC shows the lowest CPI of around 0.6 but also requires the highest total instruction count of all compilers analyzed.
It is also the only compiler that did not manage to reduce the overall instruction count for SP compared to DP.
The deprecated ICC compiler shows by far the best results and it must be hoped that its vectorization technology will be ported to LLVM.

\section{Conclusion and Outlook}\label{sec:outlook}
This paper introduced MD-Bench, a proxy-app toolbox for performance research of \ac{MD} algorithms.
It facilitates and encourages performance-related research and provides clean implementations of state-of-the-art \ac{MD} optimization schemes such as Verlet List and GROMACS MxN.
We listed and described the most important MD-Bench features and its differences to alternative proxy-apps, highlighting its use for low-level code analysis and investigation of performance implications through profiling with \ac{HPM}.

In five use cases we showed how MD-Bench can be employed for a systematic and in-depth performance analysis.
We emphasize the combination of profiling, static assembly code analysis and \ac{HPM} event measurements to gain the necessary insights for a sound performance evaluation.
We demonstrated that these techniques allow to gain an in-depth understanding of the performance and performance differences of various \ac{MD} algorithms on multiple architectures using different compilers.
The results from the analysis we conducted are snapshots representing the current state of \ac{MD} algorithms, compiler capabilities, and processor features.
With MD-Bench it is possible to repeat these studies in a reproducible and structured manner, following and documenting future developments in compiler and processor technology.
We hope that MD-Bench can also serve as a central instance of \ac{MD} algorithm reference implementations that foster the understanding and improvement of those algorithms and enable further research for \ac{MD} performance optimization.

MD-Bench is mature and usable with support for CPU with OpenMP and GPU targets.
There are still important open points to cover in future work.
We want to consider more optimized \ac{MD} schemes such as, e.g., found in NAMD.
A competitive distributed-memory parallelization with MPI will be started shortly.
Another work in progress is the implementation of specific optimization schemes for GPU accelerators (e.g. GROMACS super-cluster strategy).
Performance engineering on GPUs with MD-Bench is an important research focus for the near future.

A project like MD-Bench is an ongoing effort, keeping track with recent developments and supporting novel hardware architectures.
With this work we hope to encourage others to participate and contribute to the development of MD-Bench. 

\subsubsection*{Acknowledgments}

The authors gratefully acknowledge the scientific support and HPC resources provided by the Erlangen National High Performance Computing Center (NHR@FAU) of the Friedrich-Alexander-Universität Erlangen-Nürnberg (FAU).
NHR funding is provided by federal and Bavarian state authorities.
NHR@FAU hardware is partially funded by the German Research Foundation (DFG) -- 440719683.






\end{document}